%% file: access.tex
\let\TeXyear\year
\documentclass{ieeeaccess}
\let\IEEEAccessYear\year
\let\year\TeXyear

\usepackage{cite}

\usepackage{amsmath,amssymb,amsfonts,bm,mathtools}
\usepackage{graphicx}
\usepackage{longtable}
\usepackage{supertabular}
\usepackage{textcomp}
\usepackage{booktabs}
\usepackage{array}
\usepackage{placeins}
\usepackage{adjustbox}
\usepackage{multirow}
\usepackage{url}
\usepackage{algorithm}
\usepackage{algpseudocode}
\usepackage{siunitx}
\usepackage{placeins}
\usepackage{xcolor}
\usepackage[caption=false,font=footnotesize]{subfig}
\usepackage{tikz}
\usetikzlibrary{arrows.meta,positioning,calc,shapes.geometric}
\usepackage[
    colorlinks=true,
    linkcolor=blue,
    citecolor=blue,
    urlcolor=blue
]{hyperref}
\usepackage[
    capitalise,
    noabbrev,
    nameinlink
]{cleveref}
\let\originalciteleft\citeleft
\let\originalciteright\citeright
\let\originalcitepunct\citepunct
\let\originalcitemid\citemid
\let\originalcitedash\citedash

\makeatletter
\newenvironment{breakablealgorithm}
{%
    \par\addvspace{\topsep}%
    \refstepcounter{algorithm}%
    \hrule height .8pt depth 0pt
    \kern 2pt
    \renewcommand{\caption}[2][\relax]{%
        \textbf{Algorithm~\thealgorithm.} ##2\par
        \ifx\relax##1\relax
            \addcontentsline{loa}{algorithm}
            {\protect\numberline{\thealgorithm}##2}%
        \else
            \addcontentsline{loa}{algorithm}
            {\protect\numberline{\thealgorithm}##1}%
        \fi
        \kern 2pt
        \hrule
        \kern 3pt
    }%
}
{%
    \kern 2pt
    \hrule
    \par\addvspace{\topsep}%
}
\makeatother

\renewcommand{\citeleft}{{\color{blue}\originalciteleft}}
\renewcommand{\citeright}{{\color{blue}\originalciteright}}
\renewcommand{\citepunct}{{\color{blue}\originalcitepunct}}
\renewcommand{\citemid}{{\color{blue}\originalcitemid}}
\renewcommand{\citedash}{{\color{blue}\originalcitedash}}

\renewcommand{\eqref}[1]{%
    \textcolor{blue}{\textup{(\ref{#1})}}%
}
\let\year\IEEEAccessYear
\definecolor{accessblue}{cmyk}{1,0.3,0,0.2}
\definecolor{greycolor}{cmyk}{0,0,0,0.8}
\definecolor{grey}{cmyk}{0,0,0,0.1}
\definecolor{black}{cmyk}{0,0,0,1}
\graphicspath{{figures/}}
\makeatletter
\def\xfigwd{\columnwidth}
\makeatother
\def\BibTeX{{\rm B\kern-.05em{\sc i\kern-.025em b}\kern-.08em
T\kern-.1667em\lower.7ex\hbox{E}\kern-.125emX}}
\newcommand{\mc}{\mathcal}
\newcommand{\dt}{\Delta t}
\newcommand{\dx}{\Delta x}
\newcommand{\HDV}{\mathrm{HDV}}
\newcommand{\CAV}{\mathrm{CAV}}
\newcommand{\ff}{\mathrm{ff}}
\newcommand{\vehlen}{L_{\mathrm v}}
\newcommand{\tail}{\operatorname{tail}}
\newcommand{\head}{\operatorname{head}}

\begin{document}
\history{Date of publication xxxx 00, 0000, date of current version xxxx 00, 0000.}
\doi{10.1101/ACCESS.XXXX.DOI}

\title{MIXSim: A multiclass first order traffic model with composition-dependent dynamics for mixed-autonomy urban networks}

\author{\uppercase{Muhammad Haris}\authorrefmark{1} and \uppercase{Claudio Roncoli}\authorrefmark{2,1}}
\address[1]{Department of Built Environment, Aalto University, Finland (email: muhammad.haris@aalto.fi)}
\address[2]{Mobility and Industrial Management, KU Leuven, Belgium (email: claudio.roncoli@kuleuven.be)}
\tfootnote{This work was partly supported by the Research Council of Finland through grant ALCOSTO (349327). This work has been submitted to IEEE Access for possible publication.}

\markboth{Haris and Roncoli: MIXSim: A multiclass first order traffic model for mixed-autonomy urban networks}
{Haris and Roncoli: MIXSim: A multiclass first order traffic model for mixed-autonomy urban networks}

\corresp{Corresponding author: Muhammad Haris (email: muhammad.haris@aalto.fi).}

\begin{abstract}
This paper presents a novel multiclass first order macroscopic traffic model for mixed-autonomy in signalized urban networks, called MIXSim. Human driven vehicles (HDVs) and Connected and Automated Vehicles (CAVs) are represented as separate classes while origin-destination information is retained through commodity specific states. The local vehicle class composition and class specific driving behaviour characteristics determine the parameters of a composition-dependent triangular fundamental diagram. At signalized intersections, the composition of movement demand additionally determines the saturation flow rate, while the realized movement flow is constrained by demand, effective green time, and downstream receiving capacity. Numerical experiments investigate different CAV penetration levels and CAV time gaps, together with alternative signal control strategies. The numerical experiments illustrate how changes in CAV penetration and CAV time gap affect network performance under the tested conditions, and demonstrate the use of MIXSim for comparing signal control strategies across different mixed traffic regimes.
\end{abstract}

\begin{keywords}
Connected and Automated Vehicles, Macroscopic Traffic Models, Signal Control, Cell Transmission Model, Mixed Traffic, Urban Traffic Networks.
\end{keywords}

\maketitle

\section{Introduction}
\label{sec:introduction}
\PARstart{U}{rban} traffic congestion remains a persistent societal problem and a major challenge for mobility systems, as recurrent demand peaks and finite road and intersection capacities produce delays, queues, spillback, unreliable travel times, and inefficient use of the network \cite{papageorgiou2003review}. This challenge is becoming increasingly complex with the gradual introduction of Connected and Automated Vehicles (CAVs), which are expected to coexist with human driven vehicles (HDVs) during a prolonged transition period. CAVs may differ from HDVs in their car following behavior, longitudinal spacing characteristics, discharge behavior, and routing capabilities. Consequently, traffic flow characteristics and effective capacity may depend on the local vehicle composition \cite{talebpour2016influence,papamichail2019review,mohajerpoor2019mixed}. The analysis and management of mixed-autonomy urban networks therefore require traffic models capable of representing such composition-dependent dynamics rather than relying exclusively on uniform and fixed traffic flow parameters.

Traffic simulation is widely used to evaluate traffic management and control strategies before field deployment. The appropriate modeling resolution depends on the strategy being investigated, the objectives of the study, and the performance indicators of interest \cite{barcelo2010fundamentals,dadashzadeh2024evaluating}. Microscopic packages such as PTV Vissim, Aimsun, Simulation of Urban MObility~(SUMO), and CityFlow represent individual vehicle behavior and detailed controller interactions, making them particularly suitable when trajectories, lane level effects, safety, vehicle interactions, or other disaggregate measures are central \cite{fellendorf2010vissim,casas2010aimsun,lopez2018sumo,zhang2019cityflow}. When the quantities of interest are instead aggregate network measures, such as travel time, accumulation, queue discharge, spillback, throughput, or controller performance, first order macroscopic models can provide a computationally parsimonious environment for repeated network scale experiments.

The Cell Transmission Model (CTM) is an established first order macroscopic traffic flow model that describes traffic evolution through conservation equations together with sending and receiving functions \cite{daganzo1994ctm,daganzo1995ctm2}. Its discretized link representation naturally captures queue formation, propagation, and spillback, while flows across signalized junctions can be constrained by upstream demand, effective green time, movement saturation, and downstream receiving capacity. Discrete first order kinematic wave models have subsequently been extended to represent heterogeneous and multiclass traffic, such as the FASTLANE model~\cite{vanlint2008fastlane}, as well as multilane traffic with lane changing and overtaking~\cite{laval2006lane,roncoli2015multilane,carey2015lanes,roncoli2017lanecontrol,yao2017lc,subraveti2019lane,tajdari2022lanechanging}, and class specific flow interactions \cite{tiaprasert2017multiclass}. 

Mixed-autonomy traffic introduces a further source of heterogeneity because the traffic flow characteristics of a road segment may vary with the local proportion of HDVs and CAVs. Multiclass CTM formulations have therefore been developed to represent mixed human driven and automated traffic and the resulting composition-dependent traffic flow properties \cite{levin2016multiclass,qin2019ctm,roncoli2019gapregulation}. In signalized urban networks, however, vehicle composition is relevant not only to traffic propagation along links but also to the discharge capability of individual intersection movements. A network level mixed-autonomy model should therefore be able to represent composition-dependent behavior at both the link and signalized movement levels while preserving the interactions among congestion propagation, routing, and traffic signal control.

This paper presents MIXSim, a multiclass first order macroscopic model for mixed-autonomy signalized urban networks, which is designed to capture the effect of local HDV/CAV composition on traffic flow and intersection discharge characteristics while supporting network level analysis of routing and signal control strategies. The framework provides a computationally efficient environment for investigating how changes in CAV penetration and vehicle time gap characteristics propagate from local traffic conditions to network level performance. 

The main contributions of this paper are as follows:
\begin{enumerate}
\item The formulation of a multiclass first order macroscopic traffic model for mixed-autonomy signalized urban networks that retains OD information while representing HDVs and CAVs separately throughout the network.

\item A composition-dependent traffic flow representation in which local HDV/CAV composition influences both link level traffic flow properties and movement specific discharge capability at signalized intersections.

\item An integrated network simulation framework for commodity-based routing, signalized node flows, external demand, and alternative signal-control strategies.
\end{enumerate}

\section{Related Work and Positioning}
\label{sec:related}
Traffic simulation models span microscopic, mesoscopic, and macroscopic levels of representation, with the appropriate resolution depending on the phenomena and performance measures of interest \cite{barcelo2010fundamentals,dadashzadeh2024evaluating}. Microscopic models resolve individual vehicles and are suitable when trajectories, lane changing behavior, vehicle interactions, or detailed controller responses need to be represented explicitly. Mesoscopic and macroscopic models aggregate part or all of this vehicle level behavior and can offer lower computational cost and a more compact representation when the focus is on aggregate traffic dynamics, including queue propagation, spillback, accumulation, throughput, and network control performance. At a more aggregate network scale, Network Macroscopic Fundamental Diagram (NMFD) approaches relate network accumulation to production and provide a complementary representation of urban traffic performance \cite{geroliminis2007macroscopic,geroliminis2008existence}. Recent macroscopic simulation tools such as UXsim~\cite{seo2023uxsim,seo2025uxsim} and CTM environments such as LightSim~\cite{su2026lightsim} further illustrate the usefulness of computationally efficient frameworks for network traffic analysis and reproducible control experiments.

An important line of research has extended macroscopic traffic models to mixed traffic containing human driven and automated vehicles. A multiclass CTM formulation in~\cite{levin2016multiclass} represents human driven and autonomous vehicles as distinct classes, with vehicle composition affecting link capacity and backward wave speed, and also considers intersection control within a dynamic traffic assignment framework. Compared with the multiclass CTM formulation in~\cite{levin2016multiclass}, MIXSim represents HDV/CAV composition dependence not only in link-level traffic flow parameters but also in the saturation flow of individual signalized movements. Related mixed-autonomy studies have also investigated routing and network equilibrium when road capacity depends on the proportion and platooning capability of autonomous vehicles~\cite{lazar2021routing}. A multi commodity store and forward model for mixed-autonomy is presented in~\cite{haris&claudio2026}, incorporating dynamic saturation flow rates for joint signal control and CAV routing optimization.

Traffic assignment determines how OD demand is distributed across a network and includes deterministic, stochastic, equilibrium, and dynamic formulations~\cite{ortuzar2011modelling}. Routing is particularly relevant in mixed-autonomy networks because changes in vehicle composition, capacity, and congestion may alter the relative attractiveness of alternative routes. A model intended for mixed-autonomy studies should therefore support different routing assumptions according to the purpose of the experiment while keeping the traffic loading model distinct from the selected assignment mechanism.

\section{MIXSim Modeling Framework}
\label{sec:model}
MIXSim is formulated as a multiclass first order macroscopic traffic model for an urban road network. The road network is represented by a directed graph, in which each link represents one direction of traffic movement, and each link is partitioned into spatial cells, as in the CTM. Each cell represents a finite road segment over which the traffic state is described by the number of vehicles it contains. The physical state variables therefore represent the number of vehicles of each commodity present in each cell at a given simulation time; furthermore, since external demand may not immediately enter the physical network, virtual queues are considered at the origins.

A summary of the mathematical notation used throughout the formulation is provided in Appendix~\ref{app:notation}. The following subsections define the network and cell representation, traffic flow parameters, sending and receiving functions, routing decisions, signalized node flows, and boundary terms required for the state updates.

\subsection{Network, Cells, and Commodities}
Let the traffic network be represented by the directed graph:
\begin{equation}
\mc G=(\mc V,\mc L),
\label{eq:network_graph}
\end{equation}
where $\mc V$ is the set of nodes and $\mc L$ is the set of directed links. Physically, a node represents an intersection or a network boundary point at which traffic can enter, leave, or transfer between links, while a link represents a road segment connecting two nodes in a given direction of travel. Link $\ell\in\mc L$ starts at $\tail(\ell)$ and ends at $\head(\ell)$. For each node $j\in\mc V$, the incoming and outgoing link sets are:
\begin{align}
\mc I(j)&=\{\ell\in\mc L:\head(\ell)=j\},\nonumber\\
\mc O(j)&=\{\ell\in\mc L:\tail(\ell)=j\}.
\label{eq:node_link_sets}
\end{align}

Each link $\ell$ is divided into $I_\ell$ consecutive cells indexed by $i=1,\ldots,I_\ell$ in the direction of travel. Each cell represents a finite portion of the corresponding road segment, and its length is denoted by $\dx(\ell,i)$.

The vehicle class set is:
\begin{equation}
\mc M=\{\HDV,\CAV\}.
\label{eq:vehicle_classes}
\end{equation}
A commodity $h\in\mc H$ is defined by its origin, destination, and vehicle class:
\begin{equation}
\begin{aligned}
h=\bigl(o(h),r(h),m(h)\bigr),
\qquad
&o(h),r(h)\in\mc V,\\ 
&m(h)\in\mc M.
\end{aligned}
\label{eq:commodity_definition}
\end{equation}
Thus, vehicles associated with the same OD pair but belonging to different vehicle classes are represented as different commodities.

The state $n(\ell,i,h,t)$ denotes the number of vehicles of commodity $h$ in cell $(\ell,i)$ at discrete time $t$. The aggregate and class specific vehicle counts are, respectively:
\begin{align}
n(\ell,i,t)&=\sum_{h\in\mc H}n(\ell,i,h,t),\nonumber\\
n(\ell,i,m,t)&=\sum_{\substack{h\in\mc H\\m(h)=m}}n(\ell,i,h,t).
\label{eq:aggregate_counts}
\end{align}
The aggregate traffic density is:
\begin{equation}
\rho(\ell,i,t)=\frac{n(\ell,i,t)}{\dx(\ell,i)}.
\label{eq:density}
\end{equation}

The vehicle counts constitute the conserved state of the model, while $\rho(\ell,i,t)$ is the corresponding density used to evaluate the fundamental diagram and the associated traffic flow rates.

For a nonempty cell, the local share of class $m$ is obtained directly from the state, according to:
\begin{equation}
\pi(\ell,i,m,t)=
\begin{cases}
\dfrac{n(\ell,i,m,t)}{n(\ell,i,t)}, & n(\ell,i,t)>0,\\[8pt]
\pi^0(\ell,i,m), & n(\ell,i,t)=0,
\end{cases}
\label{eq:composition}
\end{equation}
which also considers that, when a cell is empty, a prescribed reference composition $\pi^0(\ell,i,m)$ is used, satisfying:
\begin{equation}
\sum_{m\in\mc M}\pi^0(\ell,i,m)=1.
\label{eq:reference_composition}
\end{equation}

\subsection{Composition-Dependent Fundamental Diagram}
We adopt a triangular Fundamental Diagram (FD) whose parameters depend on the vehicle class composition within each cell. Let $v^{\ff}(\ell)$ denote the free flow speed on link $\ell$. Following established multiclass CTM formulations, human driven and automated vehicles are represented as distinct classes, and their proportions in cell $(\ell,i)$ influence the corresponding capacity and backward congested wave speed \cite{levin2016multiclass,qin2019ctm}.

Let $\tau(\ell,m,t)>0$ denote the time gap associated with vehicle class $m$ on link $\ell$ at time $t$, representing the desired temporal separation used to determine the equilibrium longitudinal spacing between successive vehicles. At speed $v$, the distance associated with this temporal separation is $v\tau(\ell,m,t)$. When class behavior is assumed constant in a particular experiment, $\tau(\ell,m,t)$ reduces to a class specific constant. This parameter plays the same modeling role as the class specific time gap used in mixed traffic CTM formulations such as \cite{levin2016multiclass}.

The composition weighted time gap in cell $(\ell,i)$ is:
\begin{equation}
\bar{\tau}(\ell,i,t)
=
\sum_{m\in\mc M}
\pi(\ell,i,m,t)\tau(\ell,m,t).
\label{eq:tau_bar}
\end{equation}

In the traffic flow relations below, speeds are expressed in
$\mathrm{km/h}$, densities in $\mathrm{veh/km}$, and flow rates in
$\mathrm{veh/h}$. Time gaps are therefore converted to hours and lengths
to kilometers when evaluating these relations.

Using an effective standstill spacing $\vehlen$, comprising the vehicle
length and standstill gap, the critical density is defined from the
constant time gap spacing relation as:
\begin{equation}
\rho^{\mathrm{c}}(\ell,i,t)
=
\frac{1}
{v^{\ff}(\ell)\bar{\tau}(\ell,i,t)+\vehlen}.
\label{eq:critical_density}
\end{equation}

Hence, the capacity flow rate varies with the vehicle class composition through $\bar{\tau}(\ell,i,t)$.
\begin{equation}
q^{\max}(\ell,i,t)
=
v^{\ff}(\ell)\rho^{\mathrm{c}}(\ell,i,t)
=
\frac{v^{\ff}(\ell)}
{v^{\ff}(\ell)\bar{\tau}(\ell,i,t)+\vehlen}.
\label{eq:qmax}
\end{equation}
Hence, the cell capacity varies with the vehicle class composition through $\bar{\tau}(\ell,i,t)$.

Under the adopted spacing parameterization, the vehicle spacing at standstill is $\vehlen$. The corresponding jam density is:
\begin{equation}
\rho^{\mathrm{jam}}
=
\frac{1}{\vehlen}.
\label{eq:jam_density}
\end{equation}

Let $w(\ell,i,t)>0$ denote the magnitude of the backward congested wave speed. For the triangular FD, the congested branch connects the critical state $\bigl(\rho^{\mathrm c}(\ell,i,t),q^{\max}(\ell,i,t)\bigr)$ to the jam state $\bigl(\rho^{\mathrm{jam}},0\bigr)$. Its slope therefore gives:
\begin{equation}
w(\ell,i,t)
=
\frac{q^{\max}(\ell,i,t)}
{\rho^{\mathrm{jam}}-\rho^{\mathrm c}(\ell,i,t)}
=
\frac{\vehlen}
{\bar{\tau}(\ell,i,t)}.
\label{eq:wave}
\end{equation}

The resulting triangular FD for cell $(\ell,i)$ is:
\begin{equation}
q(\ell,i,\rho,t)
=
\min
\left\{
v^{\ff}(\ell)\rho,\,
w(\ell,i,t)
\left(
\rho^{\mathrm{jam}}-\rho
\right)
\right\}.
\label{eq:fd}
\end{equation}
Equation~\eqref{eq:fd} represents a composition-dependent parameterization
of the standard triangular FD~\cite{daganzo1994ctm,daganzo1995ctm2}, with the capacity and backward congested wave speed determined by the vehicle class composition within each cell.

\begin{figure}[t]
\centering
\includegraphics[width=\linewidth]
{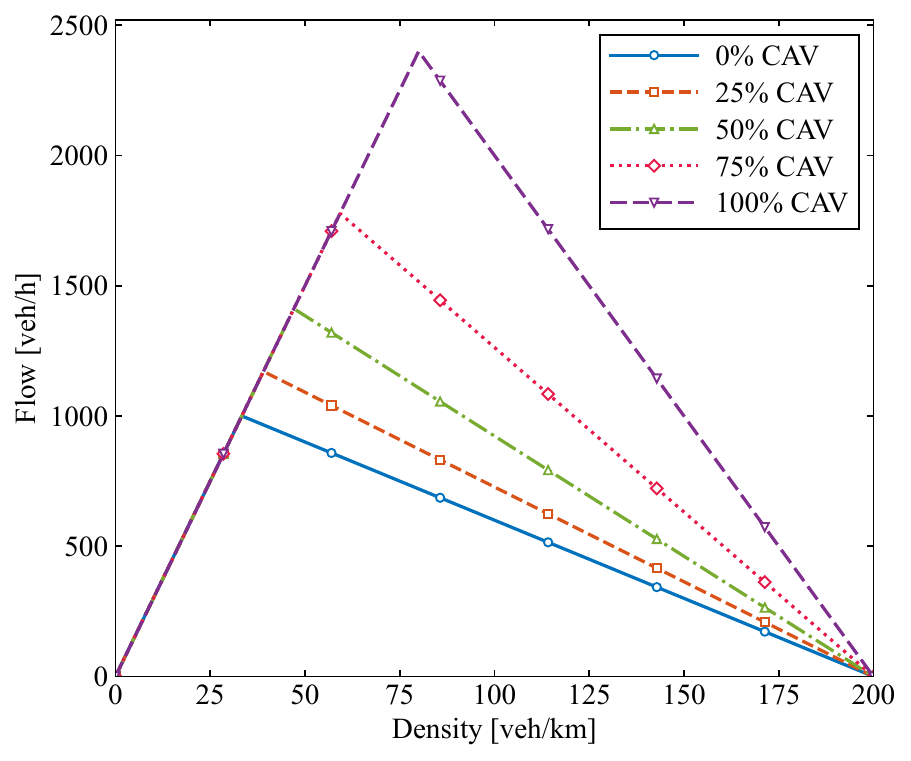}
\caption{Effect of CAV penetration on the composition-dependent triangular FD. The CAV time gap is fixed at $\tau_{\CAV}=0.9~\mathrm{s}$, and the HDV time gap at $\tau_{\HDV}=3.0~\mathrm{s}$.}
\label{fig:cav_pen}
\end{figure}

The dependence of the FD on vehicle composition is illustrated in Figure~\ref{fig:cav_pen}. For a fixed class specific time gap, increasing the CAV proportion changes the composition weighted time gap in \eqref{eq:tau_bar}. Consequently, both the capacity in \eqref{eq:qmax} and the backward congested wave speed in \eqref{eq:wave} vary with the local CAV proportion.

\begin{figure}[t]
\centering
\includegraphics[width=\linewidth]
{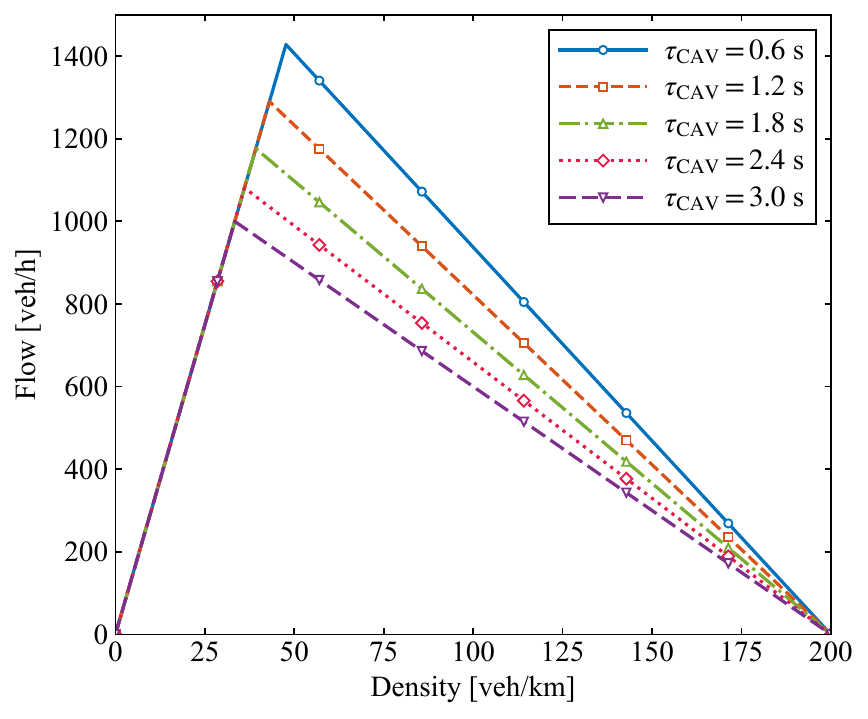}
\caption{Effect of the CAV time gap on the composition-dependent triangular FD for a fixed vehicle class composition. The CAV penetration is fixed at $45\%$.}
\label{fig:time_gap}
\end{figure}

The FD also varies with the class specific time gap even when the vehicle composition is fixed. Figure~\ref{fig:time_gap} illustrates this dependence by varying the CAV time gap while retaining a fixed CAV proportion. A smaller CAV time gap reduces $\bar{\tau}(\ell,i,t)$ and therefore increases both $q^{\max}(\ell,i,t)$ and $w(\ell,i,t)$.

These relationships show that the FD is not fixed across the network or simulation horizon. Instead, its critical state and congested branch adapt to the local vehicle class composition and the associated class specific time gaps.

Let $\dt$ denote the simulation time step in seconds. Since the backward congested wave speed increases as the time gap decreases, we define:
\begin{align}
\tau_{\min}
&=
\min_{\ell\in\mc L,\,m\in\mc M,\,t}
\tau(\ell,m,t),
\nonumber\\
w^{\max}
&=
\frac{\vehlen}
{\tau_{\min}}.
\label{eq:maximum_wave_speed}
\end{align}
The spatial and temporal discretizations are selected jointly such that:
\begin{equation}
\max
\left\{
v^{\ff}(\ell),
w^{\max}
\right\}
\frac{\dt}{3600}
\leq
\dx(\ell,i),
\label{eq:cfl}
\end{equation}
for every $\ell\in\mc L$ and $i=1,\ldots,I_\ell$.

In conventional first order traffic models, $v^{\ff}(\ell)$ is typically greater than the backward wave speed, thus the free flow propagation condition also bounds backward propagation. In the present formulation, however, $w(\ell,i,t)=\vehlen/\bar{\tau}(\ell,i,t)$ increases as the time gap decreases and can exceed the free flow speed. Both characteristic speeds are therefore considered explicitly in \eqref{eq:cfl}. For example, if $v^{\ff}=29.988~\mathrm{km/h}$, $\vehlen=0.005~\mathrm{km}$, and $\bar{\tau}=0.50~\mathrm{s}$, then \eqref{eq:wave} gives $w=36~\mathrm{km/h}$, which exceeds the free flow speed. With $\dt=5~\mathrm{s}$, the backward wave condition requires $\dx\geq0.05~\mathrm{km}$, whereas considering the free flow speed alone would require approximately $\dx\geq0.04165~\mathrm{km}$.

\subsection{Sending, Receiving, and Interior Flow}
The traffic conservation equations are discretized in space and time using a Godunov scheme for a first order traffic flow model~\cite{lebacque1996Godunov}. Under the triangular FD adopted here, the sending and receiving flow rates follow directly from the free flow and congested branches of~\eqref{eq:fd}~\cite{daganzo1994ctm,daganzo1995ctm2}. Here, the sending and receiving flow rates are evaluated using the composition-dependent values of $q^{\max}(\ell,i,t)$ and $w(\ell,i,t)$.

Commodity information is retained by distributing the aggregate flow proportionally according to the commodity composition of the upstream cell, consistent with multiclass first order traffic flow formulations~\cite{levin2016multiclass}.

The aggregate sending flow rate of cell $(\ell,i)$ is:
\begin{equation}
S(\ell,i,t)
=
\min
\left\{
v^{\ff}(\ell)\rho(\ell,i,t),
q^{\max}(\ell,i,t)
\right\}.
\label{eq:sending}
\end{equation}

The corresponding commodity specific sending flow rate is:
\begin{equation}
S(\ell,i,h,t)=
\begin{cases}
\dfrac{n(\ell,i,h,t)}{n(\ell,i,t)}\,S(\ell,i,t),
& n(\ell,i,t)>0,\\[8pt]
0,
& n(\ell,i,t)=0.
\end{cases}
\label{eq:commodity_sending}
\end{equation}

The receiving flow rate of cell $(\ell,i)$ is:
\begin{equation}
R(\ell,i,t)
=
\min
\left\{
q^{\max}(\ell,i,t),
w(\ell,i,t)
\left[
\rho^{\mathrm{jam}}
-
\rho(\ell,i,t)
\right]
\right\}.
\label{eq:receiving}
\end{equation}

For an interior boundary between cells $i$ and $i+1$ of link $\ell$,
let $Y(\ell,i,t)$ denote the aggregate traffic flow rate across the boundary. It is given by:
\begin{equation}
Y(\ell,i,t)
=
\min
\left\{
S(\ell,i,t),
R(\ell,i+1,t)
\right\},
\label{eq:interior_aggregate_flow}
\end{equation}
for $i=1,\ldots,I_\ell-1$.

The corresponding commodity specific interior flow rate is:
\begin{equation}
F(\ell,i,h,t)
=
\begin{cases}
\dfrac{n(\ell,i,h,t)}
{n(\ell,i,t)}
Y(\ell,i,t),
& n(\ell,i,t)>0,\\[8pt]
0,
& n(\ell,i,t)=0.
\end{cases}
\label{eq:interior_flow}
\end{equation} 
Thus, the aggregate flow across an interior cell boundary is first determined by the sending and receiving constraints and is then distributed among commodities according to the upstream cell composition.

\subsection{Traffic Assignment and Routing Fractions}
The network loading formulation requires routing fractions that determine how each OD commodity is distributed among the outgoing links at a node. The traffic flow model itself is independent of the particular traffic assignment or route choice method used to generate these fractions.

For commodity $h$ traveling on link $\ell$, let $\mc K(\ell,h)\subseteq\mc O(\head(\ell))$ denote the set of outgoing links from which destination $r(h)$ remains reachable. For $\head(\ell)\neq r(h)$, the routing fraction $p(\ell,k,h,t)$ denotes the fraction of commodity $h$ directed from link $\ell$ to outgoing link $k$. The routing fractions satisfy:
\begin{align}
p(\ell,k,h,t)&\geq 0,
&& k\in\mc K(\ell,h),
\nonumber\\
p(\ell,k,h,t)&=0,
&& k\notin\mc K(\ell,h),
\nonumber\\
\sum_{k\in\mc K(\ell,h)}
p(\ell,k,h,t)&=1.
\label{eq:routing_probabilities}
\end{align}

If $\head(\ell)=r(h)$, the commodity has reached its destination and no outgoing routing fraction is required. The routing fractions may be fixed or time varying and may be provided by any traffic assignment or route choice procedure that satisfies \eqref{eq:routing_probabilities}.

At each simulation step, the routing fractions are used to distribute the commodity specific sending flow rate available at the last cell of an incoming link among the admissible downstream movements. Consequently, routing determines the movement flow demand presented to the signalized node model, while the realized movement flow remains subject to effective green time, movement specific saturation, and downstream receiving capacity. The routing procedures implemented for the numerical experiments and the corresponding experiment specific configurations are described in Section~\ref{sec:sim_studies}.

\subsection{Signalized Node Flows}
The node model determines how vehicles in the last cell of an incoming link are transferred to downstream links subject to route choice, effective green time, movement discharge capability, and downstream receiving capacity.

For commodity $h$, the movement demand $D(\ell,k,h,t)$ is the flow rate available from the last cell of link $\ell$ and directed toward downstream link $k$:
\begin{equation}
D(\ell,k,h,t)\!=\!
\begin{cases}
\!p(\ell,k,h,t)S(\ell,I_\ell,h,t),
& \!\!\!\!\text{if } \head(\ell)\!\neq\! r(h),\\[5pt]
\!0,
& \text{otherwise}.
\end{cases}
\label{eq:movement_demand_commodity}
\end{equation}
The aggregate demand for movement $(\ell,k)$ is:
\begin{equation}
D(\ell,k,t)=\sum_{h\in\mc H}D(\ell,k,h,t).
\label{eq:movement_demand}
\end{equation}
Thus, $D(\ell,k,h,t)$ is the commodity specific turning flow demand, while $D(\ell,k,t)$ is the total flow demand for movement $(\ell,k)$ during the current simulation step.

To determine the class composition of the movement, define:
\begin{equation}
D^{m}(\ell,k,t)=
\sum_{\substack{h\in\mc H\\m(h)=m}}D(\ell,k,h,t),
\qquad m\in\mc M.
\label{eq:movement_class_demand}
\end{equation}
The movement specific class share is:
\begin{equation}
\alpha(\ell,k,m,t)=
\begin{cases}
\dfrac{D^{m}(\ell,k,t)}{D(\ell,k,t)}, & D(\ell,k,t)>0,\\[10pt]
\pi(\ell,I_\ell,m,t), & D(\ell,k,t)=0.
\end{cases}
\label{eq:movement_class_share}
\end{equation}
When no vehicle currently requests the movement, the composition of the upstream last cell is used as the reference composition.

The composition-weighted time gap parameter associated with movement $(\ell,k)$ is:
\begin{equation}
\bar\tau^{\mathrm{mov}}(\ell,k,t)=
\sum_{m\in\mc M}\alpha(\ell,k,m,t)\tau(\ell,m,t).
\label{eq:movement_tau}
\end{equation}
We apply the same composition-dependent spacing principle used at the link level to individual turning movements. The movement saturation flow rate is:
\begin{equation}
\mu(\ell,k,t)=
\frac{v^{\ff}(\ell)}{v^{\ff}(\ell)\bar\tau^{\mathrm{mov}}(\ell,k,t)+\vehlen}.
\label{eq:movement_saturation}
\end{equation}

Let $g(\ell,k,t)\in[0,\dt]$ denote the effective green duration available
to movement $(\ell,k)$ during step $t$. For a step that is entirely green
or red:
\begin{equation}
g(\ell,k,t)=\sigma(\ell,k,t)\dt,
\qquad
\sigma(\ell,k,t)\in\{0,1\}.
\label{eq:binary_green}
\end{equation}
The candidate movement flow rate is:
\begin{equation}
\hat f(\ell,k,t)
=
\min
\left\{
D(\ell,k,t),
\mu(\ell,k,t)
\frac{g(\ell,k,t)}{\dt}
\right\}.
\label{eq:candidate_movement}
\end{equation}

\subsubsection{External Arrivals and Virtual-Origin Queues}
Let $\mc L^{\mathrm o}\subseteq\mc L$ denote the set of links through which external OD demand enters the modeled network. The quantity $a(\ell,h,t)$ denotes the external arrival flow rate of commodity $h$ at origin link $\ell$ during simulation step $t$.

The receiving capacity of the first cell of an origin link may be smaller than the external demand requesting entry during a simulation step. To preserve vehicle conservation and retain demand that cannot immediately enter the physical network, a virtual queue $b(\ell,h,t)$ is associated with each origin link and commodity, representing vehicles that are waiting to enter the modeled road network.

The total flow rate requesting entry into origin link $\ell$ during step $t$ is therefore:
\begin{equation}
\hat e(\ell,h,t)
=
\begin{cases}
a(\ell,h,t)
+
\dfrac{3600}{\dt}
b(\ell,h,t),
& \ell\in\mc L^{\mathrm o},\\[10pt]
0,
& \ell\notin\mc L^{\mathrm o}.
\end{cases}
\label{eq:candidate_entry}
\end{equation}

\subsubsection{Downstream Supply Allocation}
Several upstream movements and, when applicable, external arrivals may compete for the receiving capacity of the first cell of the same downstream link. First-order node models therefore require a rule for allocating finite downstream supply among competing demands \cite{tampere2011generic}. In this work, we use proportional allocation.

For receiving link $k$, the total candidate inflow rate is:
\begin{align}
\hat y^{\mathrm{in}}(k,t)
&=\sum_{h\in\mc H}\hat e(k,h,t)\nonumber\\
&\quad+\sum_{\ell\in\mc I(\tail(k))}\hat f(\ell,k,t),
\label{eq:candidate_inflow}
\end{align}
where the first term is nonzero only if $k\in\mc L^{\mathrm o}$ and the second term contains candidate transfers from links entering the upstream node of $k$.

The available receiving flow rate $R(k,1,t)$ is allocated through:
\begin{equation}
\chi(k,t)=
\begin{cases}
\displaystyle\min\left\{1,\frac{R(k,1,t)}{\hat y^{\mathrm{in}}(k,t)}\right\}, & \hat y^{\mathrm{in}}(k,t)>0,\\[12pt]
1, & \hat y^{\mathrm{in}}(k,t)=0.
\end{cases}
\label{eq:supply_factor}
\end{equation}
If the total candidate inflow rate does not exceed the receiving flow rate, then $\chi(k,t)=1$. Otherwise, all candidate inflows entering link $k$ are reduced by the same proportional factor.

The realized external entry and aggregate movement flow rates are:
\begin{align}
e(k,h,t)&=\chi(k,t)\hat e(k,h,t),\nonumber\\
f(\ell,k,t)&=\chi(k,t)\hat f(\ell,k,t).
\label{eq:realized_entry_movement}
\end{align}
The realized aggregate movement flow is distributed among commodities according to their movement demands:
\begin{equation}
f(\ell,k,h,t)\!\!=\!\!
\begin{cases}
\dfrac{\!D(\ell,k,h,t)}{D(\ell,k,t)}f(\ell,k,t), & \!\!\!D(\ell,k,t)>0,\\
0, & \!\!\!D(\ell,k,t)=0.
\end{cases}
\label{eq:commodity_node_flow}
\end{equation}

Consequently, the total realized inflow rate into the first cell of link $k$ cannot exceed the receiving flow rate $R(k,1,t)$. Internal movement demand that cannot cross the node remains in the last cell of its upstream link and may be served in a subsequent step. The downstream supply constraint therefore allows congestion to propagate upstream through queue growth and spillback while preserving vehicle conservation.

\subsubsection{Boundary Flows}
The upstream boundary flow rate of link $\ell$ consists of external entry, if $\ell$ is an origin link, together with flows from upstream links:
\begin{equation}
F(\ell,0,h,t)=
e(\ell,h,t)+
\sum_{\ell'\in\mc I(\tail(\ell))}f(\ell',\ell,h,t).
\label{eq:upstream_boundary_flow}
\end{equation}
By construction, $e(\ell,h,t)=0$ when $\ell\notin\mc L^{\mathrm o}$.

At the downstream boundary, vehicles either leave the modeled network at their destination or transfer to an outgoing link:
\begin{equation}
F(\ell,I_\ell,h,t)=
z(\ell,h,t)+
\sum_{k\in\mc O(\head(\ell))}f(\ell,k,h,t).
\label{eq:downstream_boundary_flow}
\end{equation}
Assuming unrestricted discharge at the destination, the exit flow rate is:
\begin{equation}
z(\ell,h,t)=
\begin{cases}
S(\ell,I_\ell,h,t), & \head(\ell)=r(h),\\[8pt]
0, & \head(\ell)\neq r(h).
\end{cases}
\label{eq:exit}
\end{equation}

\subsection{System State and Conservation Dynamics}
The system state comprises the commodity specific vehicle counts in the physical cells and the virtual origin queues. Once the interior and boundary flow rates have been determined for simulation step $t$, the corresponding vehicle amounts transferred during the step are obtained by multiplying the flow rates by $\dt/3600$, since $\dt$ is expressed in seconds and the flow rates are expressed in $\mathrm{veh/h}$. 

The physical cell states are updated according to vehicle conservation:
\begin{equation}
n(\ell,i,h,t+1)
=
n(\ell,i,h,t)
+
\frac{\dt}{3600}
\left[
\begin{aligned}
F(\ell,i-1,h,t)
\\ - F(\ell,i,h,t)
\end{aligned}
\right].
\label{eq:update}
\end{equation}
Equation~\eqref{eq:update} applies to every commodity $h\in\mc H$, link
$\ell\in\mc L$, and cell $i=1,\ldots,I_\ell$.

For each origin link and commodity, the corresponding virtual queue is
updated according to:
\begin{equation}
\begin{aligned}
b(\ell,h,t+1)
=
b(\ell,h,t)
+
\frac{\dt}{3600}
\left[
a(\ell,h,t)
-
e(\ell,h,t)
\right], \\
\qquad
\ell\in\mc L^{\mathrm o}.
\end{aligned}
\label{eq:queue}
\end{equation}
Thus, external demand that cannot enter the physical network during the current simulation step remains in the virtual queue and is available for entry in subsequent steps.

Algorithm~\ref{alg:mixsim} summarizes the complete computational sequence
of MIXSim, from initialization and evaluation of the composition-dependent
traffic flow parameters to routing, signalized node flow resolution,
downstream supply allocation, and simultaneous state updates at each
simulation step.

\begin{breakablealgorithm}
\caption{MIXSim network loading procedure}
\label{alg:mixsim}

\begingroup
\footnotesize
\raggedright
\sloppy
\begin{algorithmic}[1]

\Require Network $\mc G=(\mc V,\mc L)$; cell discretization;
commodity set $\mc H$; vehicle classes $\mc M$;
external demand; signal control; routing fractions or routing procedure;
$\dt$, $v^{\ff}$, $\vehlen$, and $\tau(\ell,m,t)$.

\State Initialize cell states $n(\ell,i,h,0)$ and virtual origin queues
$b(\ell,h,0)$.
\State Initialize routing fractions $p(\ell,k,h,0)$.
\State From \eqref{eq:aggregate_counts}, compute aggregate and class
specific vehicle counts at $t=0$.
\State Set $t\gets0$.

\While{$t<T$}

    \Statex
    \Statex \hspace{-\algorithmicindent}
    \textbf{A. Evaluate composition-dependent traffic parameters}

    \ForAll{links $\ell$ and cells $i$}
        \State From \eqref{eq:density}, compute the aggregate density.
        \State From \eqref{eq:composition}, determine the local class
        composition.
        \State From \eqref{eq:tau_bar}, compute the composition weighted
        time gap.
        \State From \eqref{eq:qmax} and \eqref{eq:wave}, evaluate the cell
        capacity flow rate and backward congested wave speed.
    \EndFor

    \Statex
    \Statex \hspace{-\algorithmicindent}
    \textbf{B. Determine interior cell flows}

    \ForAll{links $\ell$ and cells $i$}
        \State From \eqref{eq:sending} and
        \eqref{eq:commodity_sending}, compute the aggregate and commodity
        specific sending flow rates.
        \State From \eqref{eq:receiving}, compute the receiving flow rate.
    \EndFor

    \ForAll{interior cell boundaries $(\ell,i)$}
        \State From \eqref{eq:interior_aggregate_flow}, compute the
        aggregate interior flow rate.
        \State From \eqref{eq:interior_flow}, determine the commodity
        specific interior flow rates.
    \EndFor

    \Statex
    \Statex \hspace{-\algorithmicindent}
    \textbf{C. Determine routing fractions}

    \If{time varying routing is updated at time $t$}
        \State Obtain $p(\ell,k,h,t)$ from the selected routing or traffic
        assignment procedure subject to
        \eqref{eq:routing_probabilities}.
    \EndIf

    \Statex
    \Statex \hspace{-\algorithmicindent}
    \textbf{D. Determine signalized movement demand and service}

    \ForAll{admissible movements $(\ell,k)$}
        \State From \eqref{eq:movement_demand_commodity} and
        \eqref{eq:movement_demand}, compute commodity and aggregate
        movement flow demands.
        \State From \eqref{eq:movement_class_demand}, compute the class
        specific movement flow demand.
        \State From \eqref{eq:movement_class_share}, determine the
        movement class shares.
        \State From \eqref{eq:movement_tau}, compute the movement
        composition weighted time gap.
        \State From \eqref{eq:movement_saturation}, evaluate the movement
        saturation flow rate.
        \State Obtain the effective green duration from the signal control.
        \State From \eqref{eq:candidate_movement}, compute the candidate
        movement flow rate.
    \EndFor

    \Statex
    \Statex \hspace{-\algorithmicindent}
    \textbf{E. Process external arrivals and downstream supply}

    \ForAll{origin links $\ell\in\mc L^{\mathrm o}$ and relevant
    commodities $h$}
        \State Determine the prescribed external arrival flow rate
        $a(\ell,h,t)$.
        \State From \eqref{eq:candidate_entry}, compute the candidate
        external entry flow rate.
    \EndFor

    \ForAll{receiving links $k$}
        \State From \eqref{eq:candidate_inflow}, compute the total candidate
        inflow rate.
        \State From \eqref{eq:supply_factor}, determine the downstream
        supply factor.
        \State From \eqref{eq:realized_entry_movement}, compute the realized
        external entry and aggregate movement flow rates.
        \State From \eqref{eq:commodity_node_flow}, determine the commodity
        specific movement flow rates.
    \EndFor

    \Statex
    \Statex \hspace{-\algorithmicindent}
    \textbf{F. Form boundary flows}

    \ForAll{links $\ell$ and commodities $h$}
        \State From \eqref{eq:exit}, determine the destination exit flow
        rate.
        \State From \eqref{eq:upstream_boundary_flow} and
        \eqref{eq:downstream_boundary_flow}, form the upstream and
        downstream boundary flow rates.
    \EndFor

    \Statex
    \Statex \hspace{-\algorithmicindent}
    \textbf{G. Update system state}

    \State From \eqref{eq:update}, simultaneously update all physical cell
    states to time $t+1$.
    \State From \eqref{eq:queue}, simultaneously update all virtual origin
    queues to time $t+1$.
    \State From \eqref{eq:aggregate_counts}, compute aggregate and class
    specific vehicle counts at time $t+1$.
    \State $t\gets t+1$.

\EndWhile

\State \Return

\end{algorithmic}
\endgroup
\end{breakablealgorithm}

\section{Experimental setup}
\label{sec:sim_studies}

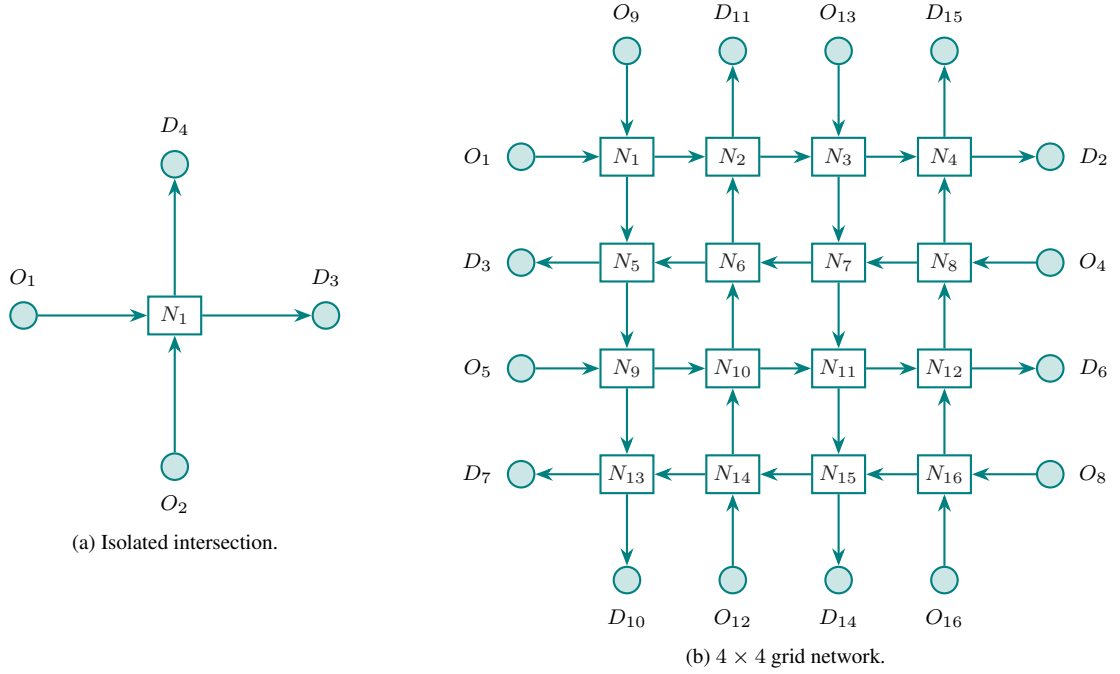
\begin{figure*}[th]
\centering
\input{figures/networks2.tex}
\caption{Networks considered in the numerical experiments:
(a) isolated intersection network with four directed links and one
intersection, and (b) \(4\times4\) grid network with 40 directed
links and 16 intersections. The symbols \(O_i\) and \(D_i\)
denote origin and destination boundary nodes, respectively.}
\label{fig:test_networks}
\end{figure*}

MIXSim is evaluated on the two signalized urban networks shown in Figure~\ref{fig:test_networks}. The first is an isolated signalized intersection with two incoming and two outgoing links. The second is a \(4\times4\) signalized grid with 16 signalized intersections, 40 directed links, eight boundary origins, and eight boundary destinations.

\subsection{Common Experimental Settings}
Unless otherwise stated, the experiments use cell length $\dx=0.1\,\mathrm{km}$, simulation step $\dt=5\,\mathrm{s}$, free flow speed $v^{\ff}=29.988\,\mathrm{km/h}$, effective standstill spacing $\vehlen=0.005\,\mathrm{km}$, HDV time gap $\tau_{\HDV}=3.0\,\mathrm{s}$, and CAV time gap $\tau_{\CAV}=0.9\,\mathrm{s}$. Time gap values are reported in seconds for readability and converted to hours when used in the traffic flow equations.

The smallest time gap considered in the experiments is
$0.60\,\mathrm{s}$. The corresponding maximum backward wave speed is:
\begin{equation}
w^{\max}
=
\frac{0.005}
{0.60/3600}
=
30\,\mathrm{km/h}.
\label{eq:experimental_max_wave_speed}
\end{equation}
Hence,
\begin{equation}
\max\{v^{\ff},w^{\max}\}
\frac{\dt}{3600}
=
30
\frac{5}{3600}
=
0.0417\,\mathrm{km}
\leq
0.1\,\mathrm{km},
\label{eq:experimental_cfl}
\end{equation}
and the CFL condition in \eqref{eq:cfl} is satisfied for all reported experiments.

For evaluating network performance, the physical network accumulation and the accumulation in virtual origin queues are defined respectively as:
\begin{align}
A(t)
&=
\sum_{\ell\in\mc L}
\sum_{i=1}^{I_\ell}
\sum_{h\in\mc H}
n(\ell,i,h,t),
\label{eq:physical_accumulation}
\\
B(t)
&=
\sum_{\ell\in\mc L^{\mathrm o}}
\sum_{h\in\mc H}
b(\ell,h,t).
\label{eq:virtual_accumulation}
\end{align}

The principal system level metric is Total Time Spent (TTS), which includes vehicles both within the physical network and in the virtual origin queues:
\begin{equation}
\mathrm{TTS}
=
\frac{\dt}{3600}
\sum_{t=0}^{T-1}
\left[A(t)+B(t)\right]
\qquad [\mathrm{veh\cdot h}].
\label{eq:tst}
\end{equation}

Peak accumulation (PA) is defined as:
\begin{equation}
\mathrm{PA}
=
\max_t
\left[A(t)+B(t)\right].
\label{eq:peak_accumulation}
\end{equation}

Clearance time (CT) is measured from the end of the demand loading period to the first simulated state satisfying:
\begin{equation}
A(t)+B(t)
\leq
10^{-4}\,\mathrm{veh}.
\label{eq:clearance_condition}
\end{equation}
The tolerance avoids interpreting negligible numerical residuals as remaining traffic. 

\subsection{Isolated Intersection Setup}
\begin{figure}[th] 
\centering 
\includegraphics[width=\linewidth]{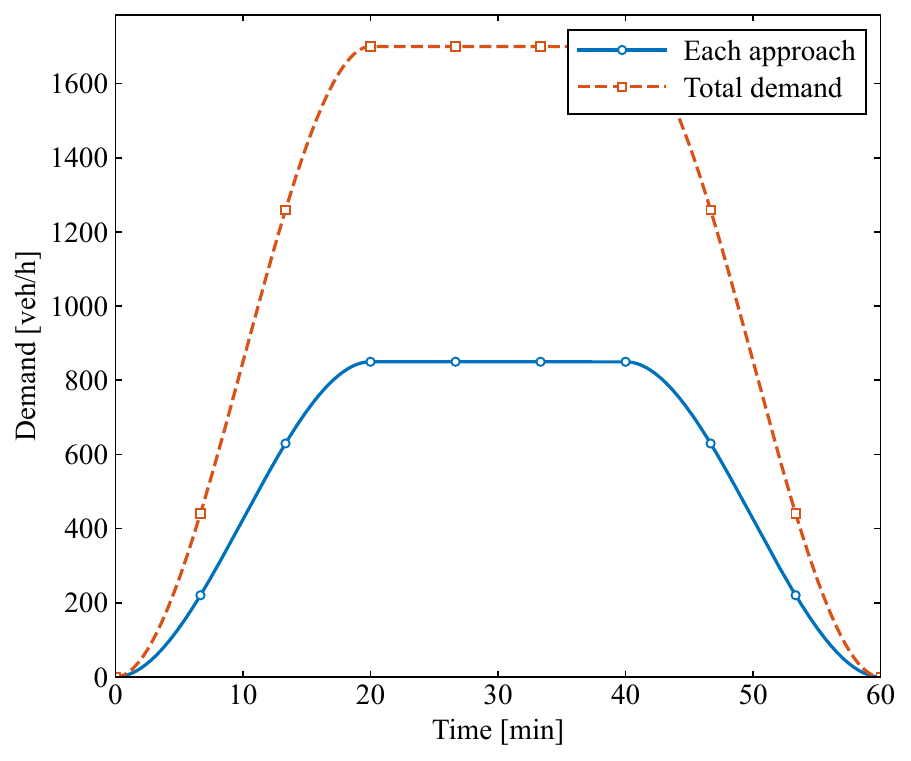} 
\caption{Demand profile used in the isolated intersection study. The two profiles correspond to the horizontal and vertical OD movements.} 
\label{fig:isolated_demand_profile} 
\end{figure}

The first experiment varies CAV penetration from 0$\%$ to 100$\%$ at the isolated intersection while keeping the signal timing and demand unchanged. Demand enters through two origin links. The intersection operates with a fixed two-phase signal plan. Each phase receives $50~\mathrm{s}$ of green, with no yellow or all-red interval, resulting in a $100~\mathrm{s}$ cycle. Each approach follows a smooth trapezoidal profile with zero initial demand, a gradual rise during 0--20~$\mathrm{min}$, a peak of \(850\,\mathrm{veh/h}\) during 20--40~$\mathrm{min}$, and a gradual decline to zero by 60~$\mathrm{min}$. The total peak demand is therefore \(1700\,\mathrm{veh/h}\), and the total requested demand over the loading period is \(1133.33\)~$\mathrm{veh}$. Figure~\ref{fig:isolated_demand_profile} shows the demand profile used for the isolated intersection.

\subsection{Grid Network Setup}
The remaining experiments are conducted on the $4\times4$ grid network. The grid uses the same temporal demand pattern as the isolated intersection, while the OD pairs and demand magnitudes are specific to the grid network. The grid demand contains eight direct OD pairs and 16 routing OD pairs. Each direct OD pair has a smooth trapezoidal profile with peak demand $580\,\mathrm{veh/h}$, while each routing OD pair has peak demand $250\,\mathrm{veh/h}$. All profiles rise smoothly from zero during 0--20~$\mathrm{min}$, remain at their peak during 20--40~$\mathrm{min}$, and fall smoothly to zero during 40--60~$\mathrm{min}$. 

\begin{figure}[th] 
\centering 
\includegraphics[width=\linewidth]{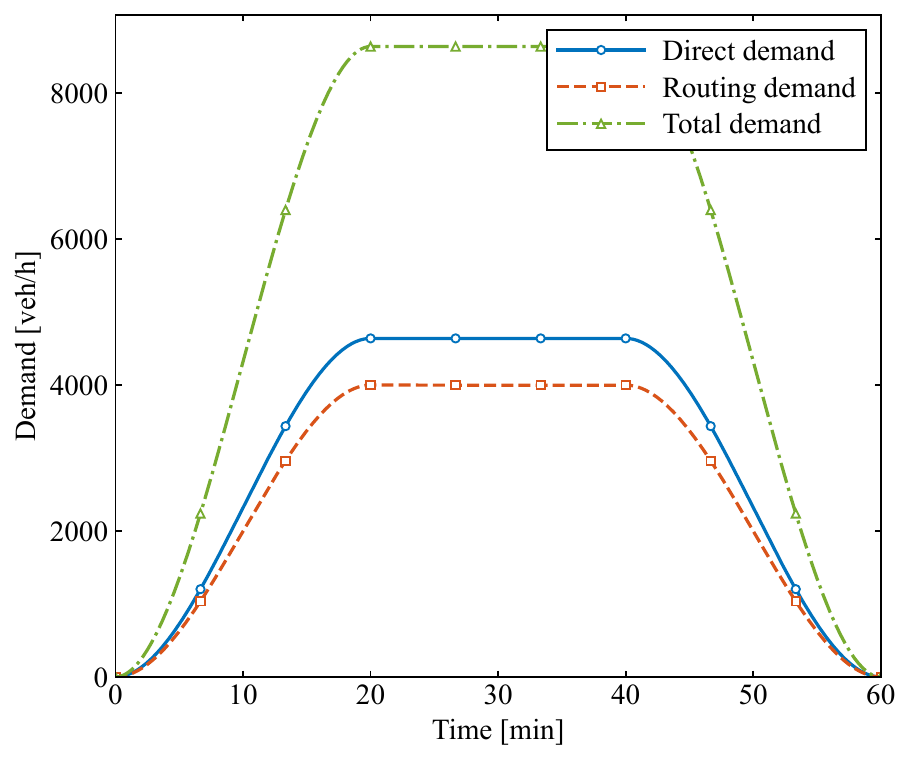} 
\caption{Demand profile used in the grid network studies. Direct demand contains the eight boundary-to-boundary OD pairs that remain on the same horizontal or vertical directions, whereas routing demand contains the 16 OD pairs that require route choice through the grid.} \label{fig:grid_demand_profile} 
\end{figure}

The resulting aggregate direct demand peak is $4640\,\mathrm{veh/h}$, the aggregate routing demand peak is $4000\,\mathrm{veh/h}$, and the total grid peak is $8640\,\mathrm{veh/h}$. The requested demand over the complete loading period is $5760~\mathrm{veh}$. Figure~\ref{fig:grid_demand_profile} shows the demand profile used in the grid network. 

Across all grid studies, the eight direct OD pairs that remain on a single horizontal or vertical corridor are assigned to fixed shortest paths based on free flow travel times.

For dynamic and stochastic assignment, routing information is updated every
$30~\mathrm{s}$ using an instantaneous travel time estimate derived from
the current traffic state. Let $u(\ell,i,t)$ denote the traffic speed implied by the fundamental diagram in \eqref{eq:fd}, obtained as $q/\rho$ for a nonempty cell and taken as $v^{\ff}(\ell)$ when the cell is empty.
\begin{equation}
\kappa(\ell,t)
=
3600
\sum_{i=1}^{I_\ell}
\frac{
\dx(\ell,i)
}{
\max\{u(\ell,i,t),\varepsilon_u\}
}.
\label{eq:routing_state_cost}
\end{equation}
To keep the routing cost finite when the estimated traffic speed approaches zero, a small positive numerical regularization parameter $\varepsilon_u$ is introduced. It only bounds the travel time used by the routing procedure and does not modify the traffic dynamics. In the numerical experiments, $\varepsilon_u=1.8~\mathrm{km/h}$ is used. Thus, $\kappa(\ell,t)$ is expressed in seconds and represents the instantaneous link travel time obtained from the current cell traffic conditions, subject to this numerical bound.

For dynamic shortest path assignment, traffic is directed toward minimum travel time downstream alternatives, with equal splitting in the case of tied minimum travel time.

For stochastic assignment, let $C(\ell,k,h,t)$ denote the estimated
remaining travel time to destination $r(h)$ when outgoing link $k$ is selected. The routing fractions are determined as: 
\begin{equation} 
\begin{aligned}
&p(\ell,k,h,t) = {} & \\
& \frac{ \exp\left[ -\theta \left( C(\ell,k,h,t)-C^{\min}(\ell,h,t) \right) \right] }{ \displaystyle \sum_{k'\in\mc K(\ell,h)} \exp\left[ -\theta \left( C(\ell,k',h,t)-C^{\min}(\ell,h,t) \right) \right] }, 
\end{aligned}
\label{eq:stochastic_routing} 
\end{equation} 
where $C^{\min}(\ell,h,t)$ is the minimum estimated remaining travel time
among the admissible alternatives. The parameter $\theta$ controls the
sensitivity of the routing fractions to differences in estimated travel
time: larger values concentrate demand more strongly on shorter travel time
alternatives, while smaller values produce a more distributed assignment.

For the CAV penetration and signal control studies, the routing OD pairs use dynamic shortest path assignment. For the CAV time gap study, stochastic assignment is used with $\theta=0.01~\mathrm{s}^{-1}$, allowing traffic to redistribute gradually among feasible alternatives rather than switching abruptly between minimum travel time routes.

For the grid CAV penetration and CAV time gap studies, signal control is held fixed using a class weighted Max Pressure controller. Phase selection is evaluated every $10~\mathrm{s}$, with a minimum green duration of $20~\mathrm{s}$ and no additional lost time when switching phases. 

The class weights used in these baseline studies are fixed as: \begin{equation} 
\beta_{\HDV} = \frac{3.0}{0.9}, 
\qquad \beta_{\CAV}=1. 
\label{eq:baseline_class_weights} 
\end{equation}
For movement $(\ell,k)$, the class weighted upstream and downstream vehicle counts are: 
\begin{align} 
Q^{\mathrm{up}}(\ell,t) &= \sum_{h\in\mc H} \beta_{m(h)} n(\ell,I_\ell,h,t), \nonumber\\ Q^{\mathrm{down}}(k,t) &= \sum_{h\in\mc H} \beta_{m(h)} n(k,1,h,t).
\label{eq:baseline_weighted_counts} 
\end{align} 
For each admissible phase $\phi$, the phase score is \begin{equation} 
\Psi_{\phi}^{\mathrm{base}}(t) = \sum_{(\ell,k)\in\mc A_{\phi}} \mu^{0}(\ell,k) \left[ Q^{\mathrm{up}}(\ell,t) - Q^{\mathrm{down}}(k,t) \right]_{+},
\label{eq:baseline_mp_score} 
\end{equation} 
where $\mc A_{\phi}$ is the set of movements served by phase $\phi$, $\mu^{0}(\ell,k)$ is the nominal saturation flow of movement $(\ell,k)$, and $[x]_{+}=\max\{x,0\}$. At each decision instant, the phase with the largest score is selected subject to the minimum green constraint.

\section{Numerical results}
\label{sec:num_results}

\subsection{Isolated Intersection}
\begin{figure}[t]
\centering
\includegraphics[width=\linewidth]{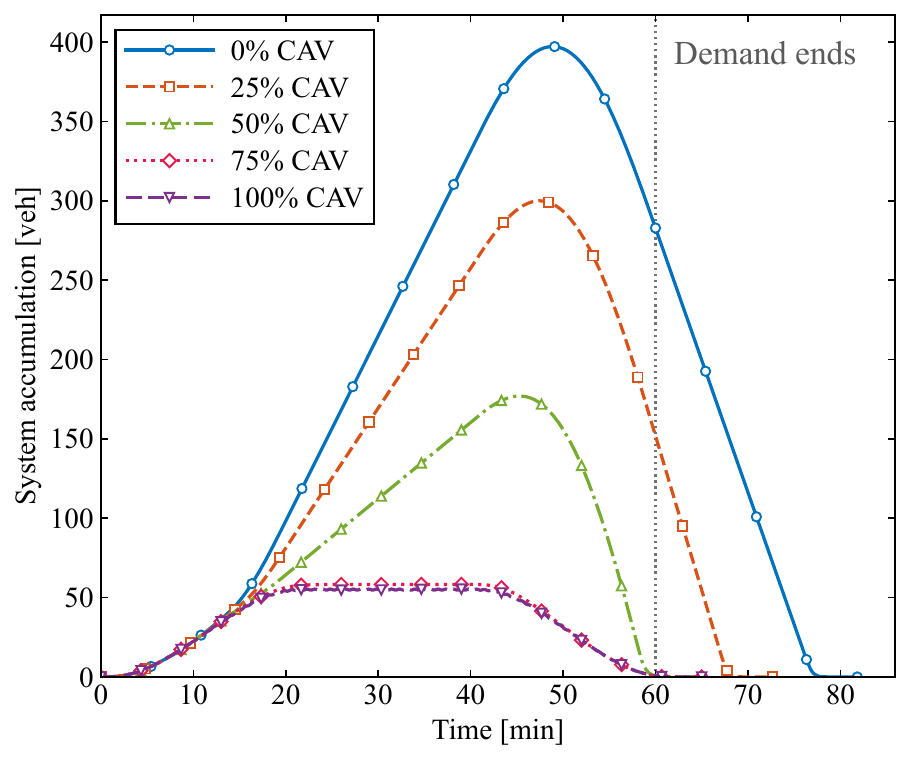}
\caption{Total system vehicles at the isolated intersection for different CAV penetration levels. The plotted total includes physical network accumulation and virtual-origin queues.}
\label{fig:isolated_net_figure}
\end{figure}

Table~\ref{tab:isolated_net_table} shows that increasing the CAV share reduces TTS, PA, and post-demand CT. The all-HDV case requires $20.42~\mathrm{min}$ after the end of the demand loading period to clear the system, whereas the cases with a CAV share of $50\%$ or higher clear within $3.58~\mathrm{min}$. Figure~\ref{fig:isolated_net_figure} further illustrates the corresponding evolution of total system accumulation over time. These improvements are consistent with the composition-dependent FD shown in Figure~\ref{fig:cav_pen}: as the CAV share increases, the composition weighted time gap decreases, thereby increasing the modeled traffic flow capacity and discharge capability.

\begin{table}[t]
\centering
\caption{CAV penetration results for the isolated intersection.}
\label{tab:isolated_net_table}
\begin{tabular}{cccc}
\toprule
\multicolumn{1}{c}{\shortstack{CAV share\\{[$\mathrm{\%}$]}}} &
\multicolumn{1}{c}{\shortstack{TTS\\{[$\mathrm{veh\cdot h}$]}}} &
\multicolumn{1}{c}{\shortstack{PA\\{[$\mathrm{veh}$]}}} &
\multicolumn{1}{c}{\shortstack{CT\\{[$\mathrm{min}$]}}} \\
\midrule
0   & 243.02 & 397.14 & 20.42 \\
25  & 162.70 & 299.95 & 11.25 \\
50  & 87.59  & 176.98 & 3.58  \\
75  & 38.00  & 58.35  & 3.58  \\
100 & 36.44  & 55.39  & 3.58  \\
\bottomrule
\end{tabular}
\end{table}

\subsection{Grid Network: CAV Penetration}
\label{subsec:grid_study1}
This experiment evaluates CAV penetration levels ranging from 0$\%$ to 100$\%$ while keeping the network configuration, demand profile, routing settings, signal control, and vehicle parameters unchanged. The corresponding performance measures are reported in Table~\ref{tab:grid_net_table}.

\begin{table}[t]
\centering
\caption{CAV penetration results for the grid network.}
\label{tab:grid_net_table}
\begin{tabular}{rrrr}
\toprule
\multicolumn{1}{c}{\shortstack{CAV share\\{[$\mathrm{\%}$]}}} &
\multicolumn{1}{c}{\shortstack{TTS\\{[$\mathrm{veh\cdot h}$]}}} &
\multicolumn{1}{c}{\shortstack{PA\\{[$\mathrm{veh}$]}}} &
\multicolumn{1}{c}{\shortstack{CT\\{[$\mathrm{min}$]}}} \\
\midrule
0   & 2114.72 & 2736.54 & 45.58 \\
25  & 1593.52 & 2324.94 & 33.67 \\
50  & 1092.78 & 1786.81 & 21.92 \\
75  & 637.53  & 1093.92 & 11.50 \\
100 & 421.01  & 636.15  & 10.92 \\
\bottomrule
\end{tabular}
\end{table}

The results show a clear reduction in TTS, PA, and post-demand CT as the CAV share increases. TTS decreases from $2114.72~\mathrm{veh\cdot h}$ in the all-HDV case to $421.01~\mathrm{veh\cdot h}$ in the all-CAV case, corresponding to a reduction of approximately $80.1\%$. Over the same range, PA decreases from $2736.54$ to $636.15~\mathrm{veh}$, while CT decreases from $45.58$ to $10.92~\mathrm{min}$.

The overall reductions in TTS, PA, and CT at higher CAV penetration are consistent with the composition-dependent fundamental diagram illustrated in Figure~\ref{fig:cav_pen}. Increasing the CAV share reduces the composition weighted time gap and increases the modeled traffic flow capacity.

\subsection{Grid Network: CAV Time Gap}
\label{subsec:grid_study3} 
This experiment examines how the CAV time gap parameter propagates through the composition-dependent traffic flow formulation. The CAV share is fixed at $45\%$, while the network, OD demand, routing configuration, signal control, and all remaining vehicle parameters are kept unchanged. The CAV time gap $\tau_{\CAV}$ is varied from $3.0~\mathrm{s}$ to $0.6~\mathrm{s}$. 

To isolate the effect of $\tau_{\CAV}$ on the traffic flow and movement saturation characteristics, the Max Pressure controller is kept unchanged across the tested CAV time gap values. The controller therefore uses the fixed baseline class weights defined in \eqref{eq:baseline_class_weights}, corresponding to the nominal values $\tau_{\HDV}=3.0~\mathrm{s}$ and $\tau_{\CAV}=0.9~\mathrm{s}$. The value of $\tau_{\CAV}$ is varied only in the traffic flow and movement saturation calculations. This avoids changing the controller objective simultaneously with the traffic model.

Table~\ref{tab:time_gap_study} summarizes the principal network level results. The quantity $\hat{\mu}_{\CAV}$ denotes the saturation flow rate obtained from \eqref{eq:movement_saturation} for a movement composed entirely of CAVs, whereas $\bar{\mu}$ denotes the mean movement saturation flow rate generated during the mixed traffic simulation. The latter depends on the time varying vehicle class composition of individual movements.

\begin{table}[t]
\centering
\caption{Varying CAV time gap results on the grid network.}
\label{tab:time_gap_study}
\begin{tabular}{cccccc}
\toprule
\multicolumn{1}{c}{$\tau_{\CAV}$} &
\multicolumn{1}{c}{$\hat{\mu}_{\CAV}$} &
\multicolumn{1}{c}{$\bar{\mu}$} &
\multicolumn{1}{c}{TTS} &
\multicolumn{1}{c}{PA} &
\multicolumn{1}{c}{CT} \\
\multicolumn{1}{c}{[s]} &
\multicolumn{1}{c}{[$\mathrm{veh/h}$]} &
\multicolumn{1}{c}{[$\mathrm{veh/h}$]} &
\multicolumn{1}{c}{[$\mathrm{veh\cdot h}$]} &
\multicolumn{1}{c}{[$\mathrm{veh}$]} &
\multicolumn{1}{c}{[$\mathrm{min}$]} \\
\midrule
3.00 &  999.93 &  999.93 & 2290.85 & 2874.38 & 78.17 \\
2.40 & 1199.90 & 1081.00 & 2019.06 & 2689.15 & 72.33 \\
1.80 & 1499.85 & 1176.33 & 1752.82 & 2477.85 & 67.08 \\
1.20 & 1999.73 & 1290.21 & 1488.53 & 2239.15 & 61.50 \\
0.60 & 2999.40 & 1428.44 & 1231.56 & 1960.85 & 56.42 \\
\bottomrule
\end{tabular}
\end{table}

\begin{figure}[t]
\centering
\includegraphics[width=\linewidth]
{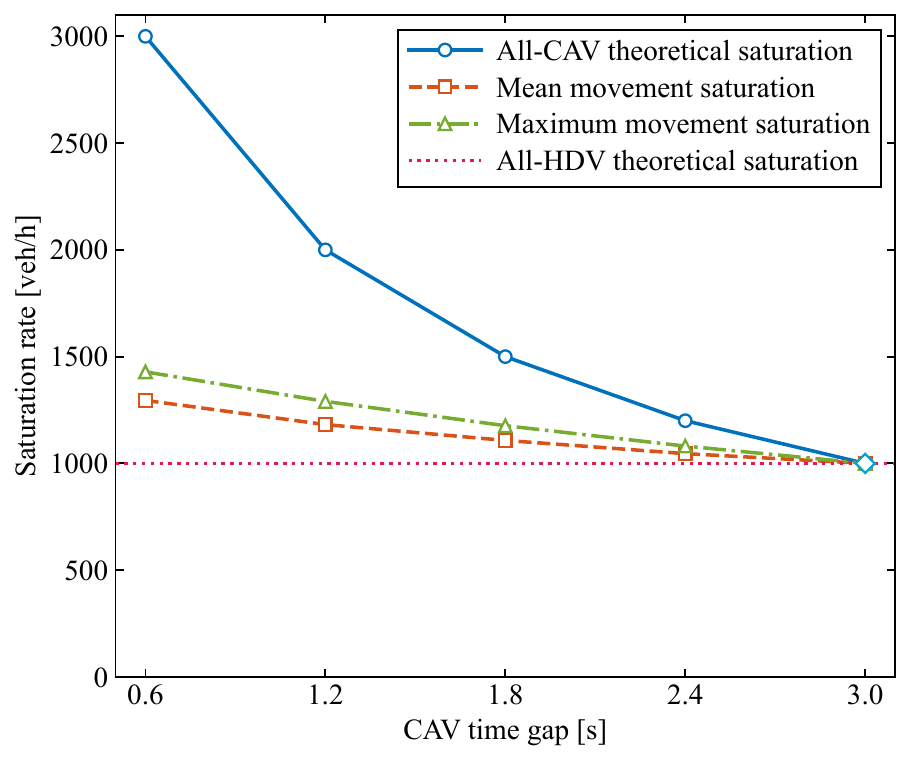}
\caption{Movement saturation rates as a function of the CAV time gap parameter for the grid network. The all-HDV theoretical saturation rate is shown as a reference.}
\label{fig:time_gap_saturation}
\end{figure}

Reducing $\tau_{\CAV}$ produces a clear increase in the available movement discharge capability. The theoretical all CAV saturation rate increases from $999.93~\mathrm{veh/h}$ at $\tau_{\CAV}=3.0\mathrm{s}$ to $2999.40~\mathrm{veh/h}$ at $\tau_{\CAV}=0.6\mathrm{s}$. The change in the actual movement saturation rate is considerably smaller because only $45\%$ of the vehicles are CAVs and the class composition varies across movements. Accordingly, the mean movement saturation rate increases from $999.93$ to $1428.44~\mathrm{veh/h}$ over the same range. This behavior is shown in Figure~\ref{fig:time_gap_saturation}, which also illustrates that the maximum movement specific saturation remains below the corresponding all CAV theoretical value whenever the movement contains both vehicle classes.

Reducing $\tau_{\CAV}$ produces a clear increase in the available movement discharge capability. The theoretical all CAV saturation rate increases from $999.93~\mathrm{veh/h}$ at $\tau_{\CAV}=3.0\mathrm{s}$ to $2999.40~\mathrm{veh/h}$ at $\tau_{\CAV}=0.6\mathrm{s}$. The change in the actual movement saturation rate is considerably smaller because only $45\%$ of the vehicles are CAVs and the class composition varies across movements. Accordingly, the mean movement saturation rate increases from $999.93$ to $1428.44~\mathrm{veh/h}$ over the same range. This behavior is shown in Figure~\ref{fig:time_gap_saturation}, which also illustrates that the maximum movement specific saturation remains below the corresponding all CAV theoretical value whenever the movement contains both vehicle classes.

Relative to the $\tau_{\CAV}=3.0\mathrm{s}$ case, reducing the time gap to $0.6\mathrm{s}$ decreases TTS from $2290.85$ to $1231.56~\mathrm{veh\cdot h}$, a reduction of approximately~$46.2\mathrm{\%}$. Peak system accumulation decreases from $2874.38$ to $1960.85~\mathrm{veh}$, while clearance time decreases from $78.17$ to $56.42~\mathrm{min}$. All five cases admit and complete the same $5760~\mathrm{veh}$, so these differences arise solely from the change in traffic evolution.

This experiment therefore demonstrates the intended coupling between the CAV time gap and network scale traffic dynamics in MIXSim. A smaller CAV time gap does not directly impose a network flow rate. Instead, it changes the composition-dependent traffic flow parameters of the cells through \eqref{eq:qmax} and \eqref{eq:wave}, as well as the saturation flow rate of individual movements through \eqref{eq:movement_saturation}. The resulting flows are then determined by the interaction among link traffic states, routing, movement demand, effective green time, and downstream receiving capacity.

Note that the tested time gap values should be interpreted as simulation parameters and not as recommended CAV operating settings. In practice, the feasible range of CAV time gaps would also depend on vehicle level safety, communication, sensing, comfort, and control requirements, which are outside the scope of these investigations.

\subsection{Grid Network: Signal Control Across Mixed-Autonomy Conditions}
\label{subsec:grid_study2}

The preceding studies show that CAV penetration and the CAV time gap modify the composition-dependent traffic characteristics represented by MIXSim. This experiment examines whether these changes also influence the relative effectiveness of signal control. Accordingly, CAV penetration, CAV time gap, and signal control strategy are varied jointly rather than comparing controllers under a single fixed mixed traffic condition.

Three signal control strategies are considered: Fixed-Time~(\emph{FT}), Demand Proportional Fixed-Time (\emph{DPFT}), and Max Pressure (\emph{MP}). \emph{FT} applies equal green splits between the two phases at each intersection using a $60~\mathrm{s}$ cycle. \emph{DPFT} also operates with a $60~\mathrm{s}$ cycle, but determines the fixed green splits from the corresponding static phase demands~\cite{webster1958traffic}. These phase demands are obtained from the prescribed OD demand using free flow shortest paths, and a minimum green duration of $10~\mathrm{s}$ is enforced.

\emph{MP} is based on the standard Max Pressure controller \cite{varaiya2013maxpressure}, but is adapted here to mixed-autonomy by weighting the vehicle counts of the two classes according to their time gap characteristics. Let $\Phi_j$ denote the set of admissible phases at node $j$, and let $\mc A_{\phi}$ denote the set of movements served by phase $\phi$. The class weight is defined as:
\begin{equation} 
\beta_m = \frac{\tau_m} {\displaystyle\min_{m'\in\mc M}\tau_{m'}}. 
\label{eq:maxpressure_class_weight} 
\end{equation} 
The corresponding class weighted upstream and downstream vehicle counts for movement $(\ell,k)$ are:
\begin{align} 
Q^{\mathrm{up}}(\ell,t) 
&= \sum_{h\in\mc H} \beta_{m(h)} n(\ell,I_\ell,h,t), 
\label{eq:maxpressure_upstream_queue} 
\\
Q^{\mathrm{down}}(k,t) &= \sum_{h\in\mc H} \beta_{m(h)} n(k,1,h,t).
\label{eq:maxpressure_downstream_queue} 
\end{align}
For phase $\phi\in\Phi_j$, the Max Pressure score is:
\begin{equation}
\Psi^{\mathrm{MP}}_{\phi}(t)
=
\sum_{(\ell,k)\in\mc A_{\phi}}
\left[
Q^{\mathrm{up}}(\ell,t)
-
Q^{\mathrm{down}}(k,t)
\right]_+,
\label{eq:maxpressure_score}
\end{equation}
where $[x]_+=\max\{x,0\}$. 

The Max Pressure controller used here differs from the baseline controller used in Sections~\ref{subsec:grid_study1} and~\ref{subsec:grid_study3}. In those experiments, each movement pressure is multiplied by the nominal saturation flow $\mu^{0}(\ell,k)$ and the class weights are kept fixed. In this experiment, the nominal saturation flow multiplier is omitted, and the class weights are computed from the time gaps of each tested case according to \eqref{eq:maxpressure_class_weight}.

The phase selected at node $j$ is:
\begin{equation}
\phi_j^\star(t)
= \operatorname*{arg\,max}_{\phi\in\Phi_j}
\Psi^{\mathrm{MP}}_{\phi}(t).
\label{eq:maxpressure_phase}
\end{equation}
The \emph{MP} decision is evaluated every $10~\mathrm{s}$, and a minimum green duration of $20~\mathrm{s}$ is enforced before switching phases. No additional lost time is applied when phases switch.

The experiment considers CAV penetration levels of $25\%$, $50\%$, and $75\%$, together with \begin{equation} \tau_{\CAV} \in \{3.6,\;1.8,\;0.9\}\ \mathrm{s}, \label{eq:control_cav_time_gap_cases} \end{equation} while $\tau_{\HDV}=3.0~\mathrm{s}$. These three values are selected to represent qualitatively different mixed traffic conditions: a CAV time gap larger than the HDV time gap, an intermediate smaller value, and the nominal CAV time gap used in the baseline experiments. Unlike the preceding time gap sensitivity study, the purpose here is not to sweep the full range of CAV time gaps, but to compare the signal controllers under three representative regimes.

The same network, demand, routing configuration, and remaining model parameters defined for the grid network are used for all controller comparisons. For every combination of CAV penetration and CAV time gap, the three controllers therefore operate under the same traffic conditions.

\begin{table*}[t]
\centering
\caption{Signal-control performance for different CAV penetration and
CAV time-gap regimes on the $4\times4$ grid. TTS reduction is evaluated
relative to FT under the same mixed-traffic conditions; ``--'' denotes
the reference FT case.}
\label{tab:control_interaction_full}
\begin{tabular}{ccccccc}
\toprule
\multicolumn{1}{c}{\shortstack{$\tau_{\CAV}$\\{[$\mathrm{s}$]}}} &
\multicolumn{1}{c}{\shortstack{CAV share\\{[$\mathrm{\%}$]}}} &
\multicolumn{1}{c}{Controller} &
\multicolumn{1}{c}{\shortstack{TTS\\{[$\mathrm{veh\cdot h}$]}}} &
\multicolumn{1}{c}{\shortstack{TTS red.\\{[$\mathrm{\%}$]}}} &
\multicolumn{1}{c}{\shortstack{PA\\{[$\mathrm{veh}$]}}} &
\multicolumn{1}{c}{\shortstack{CT\\{[$\mathrm{min}$]}}} \\
\midrule
\multirow{9}{*}{3.6}
&
\multirow{3}{*}{25}
& FT   & 2460.98 & --   & 2976.43 & 46.25 \\
&
& DPFT & 2329.23 & 5.35 & 2884.35 & 45.83 \\
&
& MP   & 2267.66 & 7.86 & 2838.03 & 48.92 \\
\cmidrule(lr){2-2}\cmidrule(lr){3-7}
&
\multirow{3}{*}{50}
& FT   & 2621.38 & --   & 3067.24 & 49.67 \\
&
& DPFT & 2484.28 & 5.23 & 2977.33 & 49.25 \\
&
& MP   & 2419.26 & 7.71 & 2930.26 & 52.58 \\
\cmidrule(lr){2-2}\cmidrule(lr){3-7}
&
\multirow{3}{*}{75}
& FT   & 2782.33 & --   & 3152.08 & 53.17 \\
&
& DPFT & 2639.90 & 5.12 & 3064.68 & 52.75 \\
&
& MP   & 2574.23 & 7.48 & 3020.34 & 55.83 \\
\midrule
\multirow{9}{*}{1.8}
&
\multirow{3}{*}{25}
& FT   & 1986.03 & --   & 2664.32 & 36.00 \\
&
& DPFT & 1869.48 & 5.87 & 2563.92 & 35.50 \\
&
& MP   & 1814.86 & 8.62 & 2515.18 & 38.92 \\
\cmidrule(lr){2-2}\cmidrule(lr){3-7}
&
\multirow{3}{*}{50}
& FT   & 1675.37 & --   & 2414.77 & 29.17 \\
&
& DPFT & 1569.73 & 6.31 & 2308.83 & 28.75 \\
&
& MP   & 1518.46 & 9.37 & 2255.84 & 32.08 \\
\cmidrule(lr){2-2}\cmidrule(lr){3-7}
&
\multirow{3}{*}{75}
& FT   & 1371.64 & --    & 2123.58 & 22.50 \\
&
& DPFT & 1277.27 & 6.88  & 2011.78 & 22.17 \\
&
& MP   & 1232.61 & 10.14 & 1957.59 & 25.25 \\
\midrule
\multirow{9}{*}{0.9}
&
\multirow{3}{*}{25}
& FT   & 1752.56 & --   & 2480.91 & 30.83 \\
&
& DPFT & 1644.06 & 6.19 & 2376.29 & 30.50 \\
&
& MP   & 1593.52 & 9.07 & 2324.94 & 33.67 \\
\cmidrule(lr){2-2}\cmidrule(lr){3-7}
&
\multirow{3}{*}{50}
& FT   & 1223.03 & --    & 1959.18 & 19.17 \\
&
& DPFT & 1134.76 & 7.22  & 1844.73 & 18.83 \\
&
& MP   & 1092.78 & 10.65 & 1786.81 & 21.92 \\
\cmidrule(lr){2-2}\cmidrule(lr){3-7}
&
\multirow{3}{*}{75}
& FT   & 729.77 & --    & 1265.53 & 8.42 \\
&
& DPFT & 664.97 & 8.88  & 1145.54 & 8.25 \\
&
& MP   & 637.53 & 12.64 & 1093.92 & 11.50 \\
\bottomrule
\end{tabular}
\end{table*}

Table~\ref{tab:control_interaction_full} shows different performance trends across the tested CAV penetration and CAV time gap conditions. When $\tau_{\CAV}=3.6~\mathrm{s}$, which exceeds the HDV time gap, increasing CAV penetration deteriorates network performance under all three controllers. Under \emph{FT}, for example, TTS increases from $2460.98~\mathrm{veh\cdot h}$ at $25\%$ CAV to $2782.33~\mathrm{veh\cdot h}$ at $75\%$ CAV. Over the same range, PA increases from $2976.43$ to $3152.08~\mathrm{veh}$ and CT from $46.25$ to $53.17~\mathrm{min}$.

In contrast, when $\tau_{\CAV}=0.9~\mathrm{s}$, increasing CAV penetration produces substantial improvements. Under \emph{FT}, TTS decreases from $1752.56~\mathrm{veh\cdot h}$ at $25\%$ CAV to $729.77~\mathrm{veh\cdot h}$ at $75\%$ CAV. The intermediate case $\tau_{\CAV}=1.8~\mathrm{s}$ exhibits the same beneficial trend, although with a smaller magnitude. Under the tested conditions, the direction and magnitude of the performance change associated with increasing CAV penetration depend on the assumed CAV time gap. These trends are consistent with the composition-dependent traffic flow relationships introduced in Section~\ref{sec:model}.

For a normalized comparison of controller performance within each mixed traffic condition, the TTS reduction obtained by controller $c$ relative to \emph{FT} is defined as:
\begin{equation}
\Delta_{\mathrm{TTS}}^{c}
=
100
\frac{
\mathrm{TTS}^{\mathrm{FT}}
-
\mathrm{TTS}^{c}
}{
\mathrm{TTS}^{\mathrm{FT}}
}.
\label{eq:control_tst_reduction}
\end{equation}
A positive value indicates lower TTS than \emph{FT} under the same tested
mixed traffic condition.

Both \emph{DPFT} and \emph{MP} reduce TTS relative to \emph{FT} in every tested mixed traffic regime, but the magnitude of the benefit depends on the CAV characteristics. For \emph{DPFT}, the TTS reduction remains approximately $5.1$--$5.4\%$ when $\tau_{\CAV}=3.6~\mathrm{s}$. At $\tau_{\CAV}=1.8~\mathrm{s}$, it increases from $5.87\%$ at $25\%$ CAV to $6.88\%$ at $75\%$ CAV, while for $\tau_{\CAV}=0.9~\mathrm{s}$ it increases from $6.19\%$ to $8.88\%$.

The interaction is stronger for \emph{MP}. When $\tau_{\CAV}=3.6~\mathrm{s}$, the TTS reduction varies only slightly from $7.86\%$ at $25\%$ CAV to $7.48\%$ at $75\%$ CAV. For $\tau_{\CAV}=1.8~\mathrm{s}$, the reduction increases from $8.62\%$ to $10.14\%$, while for $\tau_{\CAV}=0.9~\mathrm{s}$ it increases from $9.07\%$ to $12.64\%$. The relative TTS reductions therefore vary across the tested CAV penetration and time gap conditions, although the controller ordering remains unchanged.

Across all nine tested mixed traffic conditions, \emph{MP} produces the lowest TTS and PA among the three controllers, while \emph{DPFT} consistently improves these metrics relative to \emph{FT}. The ordering of the controllers with respect to TTS and PA is therefore unchanged across the tested mixed traffic conditions. Clearance time exhibits a different pattern: \emph{DPFT} produces a slightly shorter CT than \emph{FT} in every case, whereas \emph{MP} produces a longer CT despite its lower TTS and PA. For example, at $75\%$ CAV and $\tau_{\CAV}=0.9~\mathrm{s}$, \emph{MP} reduces TTS from $729.77$ to $637.53~\mathrm{veh\cdot h}$ and PA from $1265.53$ to $1093.92~\mathrm{veh}$, while CT increases from $8.42$ to $11.50~\mathrm{min}$. Thus, under the tested conditions, lower cumulative network congestion does not necessarily coincide with a shorter final clearance period.

\section{Conclusion}
\label{sec:conclusion}
This paper presented MIXSim, a multiclass first order macroscopic traffic model for mixed-autonomy in signalized urban networks. MIXSim retains OD and vehicle class information through commodity specific cell states and represents HDVs and CAVs separately throughout the network. Its main feature is a composition-dependent traffic flow representation in which the local vehicle class composition and class specific time gaps determine the parameters of the triangular fundamental diagram. The same principle is extended to signalized movements, where the composition of the traffic requesting a movement determines its saturation flow rate.

Numerical experiments conducted on an isolated intersection and a grid network demonstrate that MIXSim responds consistently to changes in vehicle class composition and CAV time gap through the resulting changes in capacity, backward congested wave speed, and movement saturation flow. The signal control experiments further demonstrate that the same traffic model can be used to compare different control strategies under several mixed traffic conditions. More extensive experiments covering different demand levels, demand patterns, network configurations, routing conditions, and model parameters would be required before drawing broader conclusions regarding the traffic performance effects of CAV penetration or particular signal control strategies.

MIXSim is intentionally macroscopic and is therefore not intended to reproduce lane level interactions, individual vehicle trajectories, explicit platooning, communication processes, or vehicle level safety constraints. These phenomena are more appropriately represented using microscopic models. Future work will focus on calibration and validation against microscopic simulation and empirical traffic data, extensions of the node model to represent more detailed movement and lane group interactions, and evaluation on larger and more heterogeneous urban networks. The model may also provide a basis for developing and testing traffic control and routing methods that explicitly account for spatially and temporally varying vehicle class composition and CAV traffic characteristics.

\section*{Software Availability}
The source code of MIXSim, including the model, the two-intersection tutorial, scripts for reproducing the numerical experiments, and automated tests, is publicly available at \url{https://github.com/HarisFateh/MIXSim}.

\appendices
\section{}
\label{app:notation}
\input{tables/math_not.tex}
\clearpage

\bibliographystyle{IEEEtran}
\bibliography{references}

\phantomsection
\begin{IEEEbiography}[{\includegraphics[width=1in,height=1.25in,clip,keepaspectratio]{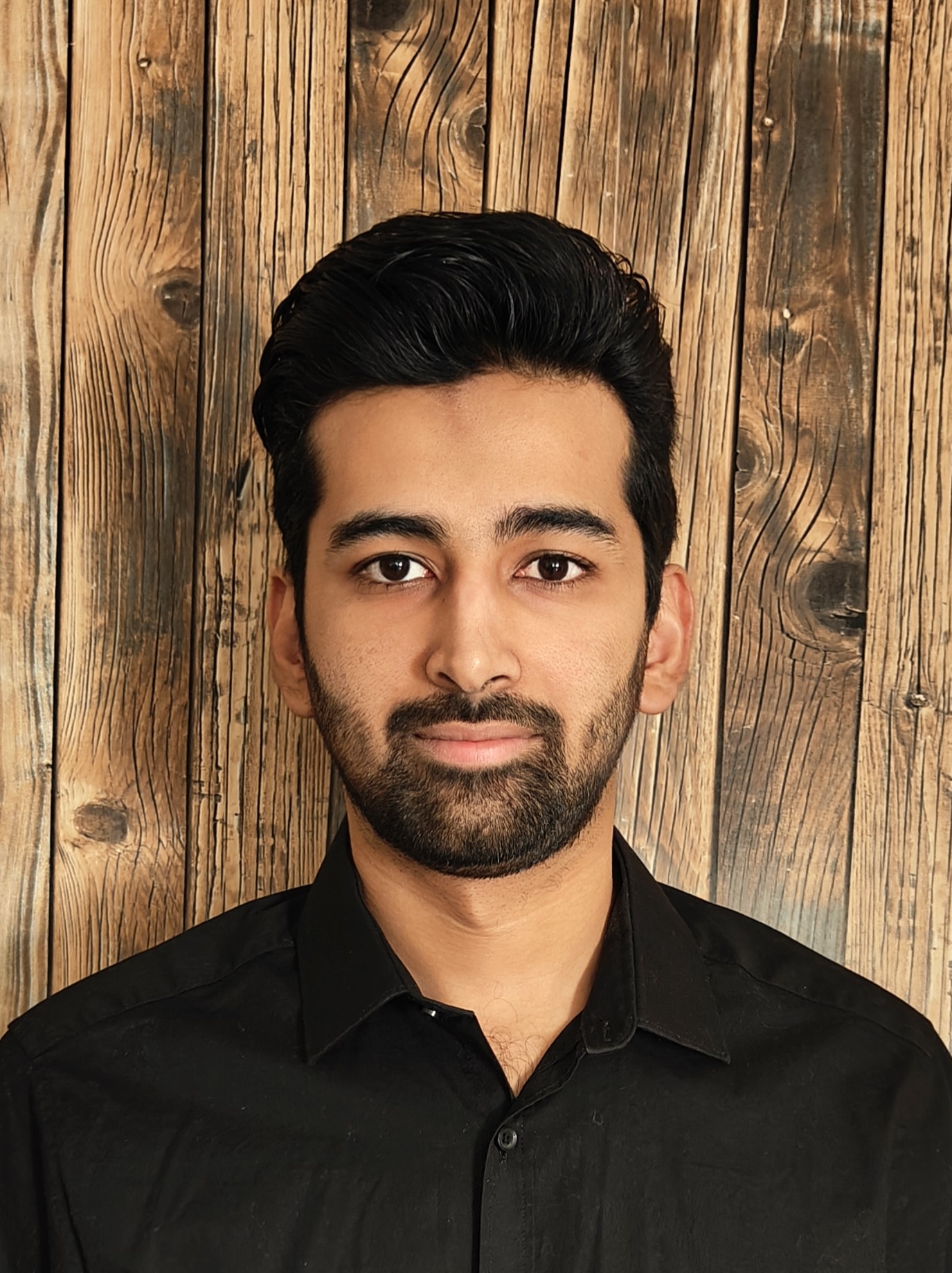}}]{Muhammad Haris } is a Doctoral Researcher at the Department of Built Environment, Aalto University, Finland, where he is pursuing a Doctor of Science (Technology) in the School of Engineering. He previously obtained his Master’s degree in Electrical and Computer Engineering from King Abdullah University of Science \& Technology, Saudi Arabia, and his Bachelor’s degree in the same field from the National University of Sciences \& Technology, Pakistan.
Prior to his doctoral studies, he worked as a Graduate Research Assistant at Robotics, Intelligent Systems, and Control Lab, King Abdullah University of Science \& Technology.
His current work focuses on developing model-based solutions for complex urban mobility challenges. His broader interests include real-time traffic management systems, air traffic management, applied optimization, mathematical modeling, and intelligent transport systems.
\end{IEEEbiography}

\begin{IEEEbiography}[{\includegraphics[width=1in,height=1.25in,clip,keepaspectratio]{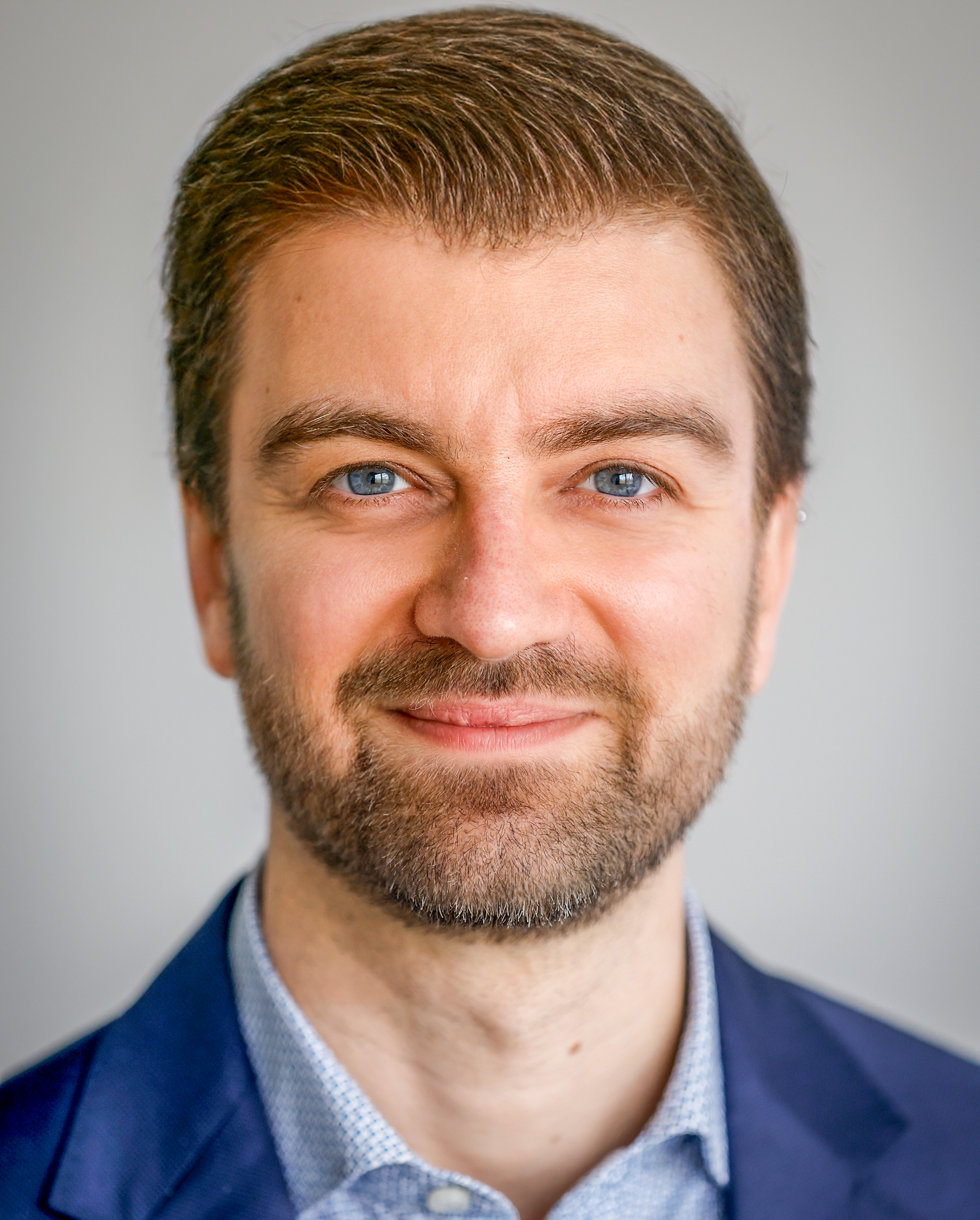}}]{Claudio Roncoli} (Senior Member, IEEE) is an Associate Professor at the Division of Mobility and Industrial Management, KU Leuven, Belgium. Before joining KU Leuven, he was an Associate Professor at Aalto University, Finland, where he still holds a Visiting Professor position. Previously, he was a researcher at the Technical University of Crete, Greece, and the University of Genova, Italy, as well as a visiting researcher at the University of Queensland and the University of Sydney, Australia, and at the Imperial College London, UK. 
Claudio has been involved in several national and international research projects as a principal investigator. His research interests include real-time traffic management; modeling, optimisation, and control of traffic systems with connected and automated vehicles; as well as smart mobility and intelligent transport systems.
\end{IEEEbiography}

\EOD

\vfill

\end{document}

%% file: figures/networks2.tex

\tikzset{
    >=Stealth,
    intersection/.style={
        rectangle,
        draw=teal,
        thick,
        minimum width=0.70cm,
        minimum height=0.50cm,
        inner sep=1pt,
        text=black,
        font=\footnotesize
    },
    boundary/.style={
        circle,
        draw=teal,
        thick,
        fill=teal!20,
        minimum size=3.5mm,
        inner sep=0pt
    },
    link/.style={
        ->,
        thick,
        teal
    }
}

\subfloat[Isolated intersection.\label{fig:isolated_network}]{%
\begin{minipage}[c]{0.35\textwidth}
\centering
\begin{adjustbox}{max width=\linewidth}
\begin{tikzpicture}[every node/.append style={font=\footnotesize}]
\node[intersection] (N1) at (0,0) {$N_1$};
\node[boundary] (O1) at (-2.0,0) {};
\node[above=2pt of O1] {$O_1$};
\node[boundary] (O2) at (0,-2.0) {};
\node[below=2pt of O2] {$O_2$};
\node[boundary] (D3) at (2.0,0) {};
\node[above=2pt of D3] {$D_3$};
\node[boundary] (D4) at (0,2.0) {};
\node[above=2pt of D4] {$D_4$};
\draw[link] (O1) -- (N1) node[midway, above] {} ;
\draw[link] (O2) -- (N1) node[midway, left]{};
\draw[link] (N1) -- (D3) node[midway, above] {};
\draw[link] (N1) -- (D4) node[midway, right] {};
\end{tikzpicture}
\end{adjustbox}
\end{minipage}}%
\hspace{0.02\textwidth}%
\subfloat[$4\times4$ grid network.\label{fig:grid_network}]{%
\begin{minipage}[c]{0.52\textwidth}
\centering
\begin{adjustbox}{max width=\linewidth}
\begin{tikzpicture}[every node/.append style={font=\footnotesize}]
\node[intersection] (N1)  at (0,0)     {$N_1$};
\node[intersection] (N2)  at (1.4,0)   {$N_2$};
\node[intersection] (N3)  at (2.8,0)   {$N_3$};
\node[intersection] (N4)  at (4.2,0)   {$N_4$};
\node[intersection] (N5)  at (0,-1.4)   {$N_5$};
\node[intersection] (N6)  at (1.4,-1.4) {$N_6$};
\node[intersection] (N7)  at (2.8,-1.4) {$N_7$};
\node[intersection] (N8)  at (4.2,-1.4) {$N_8$};
\node[intersection] (N9)  at (0,-2.8)   {$N_9$};
\node[intersection] (N10) at (1.4,-2.8) {$N_{10}$};
\node[intersection] (N11) at (2.8,-2.8) {$N_{11}$};
\node[intersection] (N12) at (4.2,-2.8) {$N_{12}$};
\node[intersection] (N13) at (0,-4.2)   {$N_{13}$};
\node[intersection] (N14) at (1.4,-4.2) {$N_{14}$};
\node[intersection] (N15) at (2.8,-4.2) {$N_{15}$};
\node[intersection] (N16) at (4.2,-4.2) {$N_{16}$};
\node[boundary] (O1) at (-1.4,0) {}; \node[left=2pt of O1] {$O_1$};
\node[boundary] (D2) at (5.6,0) {}; \node[right=2pt of D2] {$D_2$};
\node[boundary] (D3) at (-1.4,-1.4) {}; \node[left=2pt of D3] {$D_3$};
\node[boundary] (O4) at (5.6,-1.4) {}; \node[right=2pt of O4] {$O_4$};
\node[boundary] (O5) at (-1.4,-2.8) {}; \node[left=2pt of O5] {$O_5$};
\node[boundary] (D6) at (5.6,-2.8) {}; \node[right=2pt of D6] {$D_6$};
\node[boundary] (D7) at (-1.4,-4.2) {}; \node[left=2pt of D7] {$D_7$};
\node[boundary] (O8) at (5.6,-4.2) {}; \node[right=2pt of O8] {$O_8$};
\node[boundary] (O9) at (0,1.4) {}; \node[above=2pt of O9] {$O_9$};
\node[boundary] (D11) at (1.4,1.4) {}; \node[above=2pt of D11] {$D_{11}$};
\node[boundary] (O13) at (2.8,1.4) {}; \node[above=2pt of O13] {$O_{13}$};
\node[boundary] (D15) at (4.2,1.4) {}; \node[above=2pt of D15] {$D_{15}$};
\node[boundary] (D10) at (0,-5.6) {}; \node[below=2pt of D10] {$D_{10}$};
\node[boundary] (O12) at (1.4,-5.6) {}; \node[below=2pt of O12] {$O_{12}$};
\node[boundary] (D14) at (2.8,-5.6) {}; \node[below=2pt of D14] {$D_{14}$};
\node[boundary] (O16) at (4.2,-5.6) {}; \node[below=2pt of O16] {$O_{16}$};
\draw[link] (O1) -- (N1) node[midway, above] {};
\draw[link] (N1) -- (N2) node[midway, above] {};
\draw[link] (N2) -- (N3) node[midway, above] {};
\draw[link] (N3) -- (N4) node[midway, above] {};
\draw[link] (N4) -- (D2) node[midway, above] {};
\draw[link] (O4) -- (N8) node[midway, above]{} ;
\draw[link] (N8) -- (N7) node[midway, above] {};
\draw[link] (N7) -- (N6) node[midway, above] {};
\draw[link] (N6) -- (N5) node[midway, above] {};
\draw[link] (N5) -- (D3) node[midway, above] {};
\draw[link] (O5) -- (N9) node[midway, above]{} ;
\draw[link] (N9) -- (N10) node[midway, above]{} ;
\draw[link] (N10) -- (N11) node[midway, above] {};
\draw[link] (N11) -- (N12) node[midway, above] {};
\draw[link] (N12) -- (D6) node[midway, above] {};
\draw[link] (O8) -- (N16) node[midway, above] {};
\draw[link] (N16) -- (N15) node[midway, above] {};
\draw[link] (N15) -- (N14) node[midway, above] {};
\draw[link] (N14) -- (N13) node[midway, above] {};
\draw[link] (N13) -- (D7) node[midway, above] {};
\draw[link] (O9) -- (N1) node[midway, left] {};
\draw[link] (N1) -- (N5) node[midway, left] {};
\draw[link] (N5) -- (N9) node[midway, left] {};
\draw[link] (N9) -- (N13) node[midway, left] {};
\draw[link] (N13) -- (D10) node[midway, left] {};
\draw[link] (O12) -- (N14) node[midway, left] {};
\draw[link] (N14) -- (N10) node[midway, left] {};
\draw[link] (N10) -- (N6) node[midway, left] {};
\draw[link] (N6) -- (N2) node[midway, left] {};
\draw[link] (N2) -- (D11) node[midway, left] {};
\draw[link] (O13) -- (N3) node[midway, left] {};
\draw[link] (N3) -- (N7) node[midway, left] {};
\draw[link] (N7) -- (N11) node[midway, left] {};
\draw[link] (N11) -- (N15) node[midway, left] {};
\draw[link] (N15) -- (D14) node[midway, left] {};
\draw[link] (O16) -- (N16) node[midway, left] {};
\draw[link] (N16) -- (N12) node[midway, left] {};
\draw[link] (N12) -- (N8) node[midway, left] {};
\draw[link] (N8) -- (N4) node[midway, left] {};
\draw[link] (N4) -- (D15) node[midway, left] {};
\end{tikzpicture}
\end{adjustbox}
\end{minipage}}

%% file: tables/math_not.tex
\begin{table*}[!t]
\centering
\tablebodyfont
\renewcommand{\arraystretch}{1.10}
\setlength{\tabcolsep}{3pt}

\caption{Mathematical notations}
\label{tab:mathematical_notation}

\begin{minipage}[t]{0.485\textwidth}
\centering

\begin{tabular}{
>{\raggedright\arraybackslash}p{0.26\columnwidth}
>{\raggedright\arraybackslash}p{0.66\columnwidth}
}
\toprule
{\tableheadfont Symbol} &
{\tableheadfont Description} \\
\midrule

\multicolumn{2}{l}{\textbf{Network, indices, and commodities}}\\

$\mathcal{G}=(\mathcal{V},\mathcal{L})$
&
Directed traffic network.
\\

$\mathcal{V}$
&
Set of network nodes.
\\

$\mathcal{L}$
&
Set of directed network links.
\\

$j$
&
Node index.
\\

$\ell,k,\ell'$
&
Directed link indices.
\\

$\operatorname{tail}(\ell)$
&
Upstream node of link $\ell$.
\\

$\operatorname{head}(\ell)$
&
Downstream node of link $\ell$.
\\

$\mathcal{I}(j)$
&
Set of links entering node $j$.
\\

$\mathcal{O}(j)$
&
Set of links leaving node $j$.
\\

$I_{\ell}$
&
Number of cells on link $\ell$.
\\

$i$
&
Cell index, $i=1,\ldots,I_{\ell}$.
\\

$\Delta x(\ell,i)$
&
Length of cell $(\ell,i)$, expressed in $\mathrm{km}$ in the traffic
flow equations.
\\

$\Delta t$
&
Simulation time step, expressed in $\mathrm{s}$.
\\

$t$
&
Discrete simulation time index.
\\

$T$
&
Final discrete simulation time index.
\\

$\mathcal{M}$
&
Set of vehicle classes,
$\mathcal{M}=\{\mathrm{HDV},\mathrm{CAV}\}$.
\\

$m$
&
Vehicle class index.
\\

$\mathcal{H}$
&
Set of OD and vehicle class commodities.
\\

$h$
&
Commodity index.
\\

$o(h)$
&
Origin associated with commodity $h$.
\\

$r(h)$
&
Destination associated with commodity $h$.
\\

$m(h)$
&
Vehicle class associated with commodity $h$.
\\

\midrule
\multicolumn{2}{l}{\textbf{Cell states and traffic composition}}\\

$n(\ell,i,h,t)$
&
Number of vehicles of commodity $h$ in cell $(\ell,i)$ at time $t$.
\\

$n(\ell,i,t)$
&
Total number of vehicles in cell $(\ell,i)$ at time $t$.
\\

$n(\ell,i,m,t)$
&
Number of vehicles of class $m$ in cell $(\ell,i)$ at time $t$.
\\

$\rho(\ell,i,t)$
&
Aggregate traffic density in cell $(\ell,i)$ at time $t$,
expressed in $\mathrm{veh/km}$.
\\

$\pi(\ell,i,m,t)$
&
Local fraction of vehicle class $m$ in cell $(\ell,i)$ at time $t$.
\\

$\pi^{0}(\ell,i,m)$
&
Reference vehicle class composition used when cell $(\ell,i)$ is empty.
\\

\midrule
\multicolumn{2}{l}{\textbf{Composition dependent traffic flow parameters}}\\

$\tau(\ell,m,t)$
&
Time gap parameter of vehicle class $m$ on link $\ell$ at time $t$;
reported in $\mathrm{s}$ and converted to hours in the traffic flow
relations.
\\

$\bar{\tau}(\ell,i,t)$
&
Composition weighted time gap in cell $(\ell,i)$.
\\

$L_{\mathrm v}$
&
Effective standstill spacing comprising vehicle length and standstill
gap; expressed in $\mathrm{km}$ in the traffic flow equations.
\\

$v^{\mathrm{ff}}(\ell)$
&
Free flow speed on link $\ell$, expressed in $\mathrm{km/h}$.
\\

$\rho^{\mathrm{c}}(\ell,i,t)$
&
Composition dependent critical density of cell $(\ell,i)$,
expressed in $\mathrm{veh/km}$.
\\

$\rho^{\mathrm{jam}}$
&
Jam density, expressed in $\mathrm{veh/km}$.
\\

$q^{\max}(\ell,i,t)$
&
Composition dependent capacity flow rate of cell $(\ell,i)$,
expressed in $\mathrm{veh/h}$.
\\

$w(\ell,i,t)$
&
Magnitude of the backward congested wave speed, expressed in
$\mathrm{km/h}$.
\\

$\tau_{\min}$
&
Minimum vehicle time gap considered in the model.
\\

$w^{\max}$
&
Maximum backward congested wave speed magnitude.
\\

$q(\ell,i,\rho,t)$
&
Traffic flow rate given by the composition dependent triangular
fundamental diagram at density $\rho$, expressed in $\mathrm{veh/h}$.
\\

\bottomrule
\end{tabular}

\end{minipage}
\hfill
\begin{minipage}[t]{0.485\textwidth}
\centering

\begin{tabular}{
>{\raggedright\arraybackslash}p{0.26\columnwidth}
>{\raggedright\arraybackslash}p{0.66\columnwidth}
}
\toprule
{\tableheadfont Symbol} &
{\tableheadfont Description} \\
\midrule

\multicolumn{2}{l}{\textbf{Sending, receiving, and interior flow}}\\

$S(\ell,i,t)$
&
Aggregate sending flow rate of cell $(\ell,i)$,
expressed in $\mathrm{veh/h}$.
\\

$S(\ell,i,h,t)$
&
Sending flow rate of commodity $h$ from cell $(\ell,i)$,
expressed in $\mathrm{veh/h}$.
\\

$R(\ell,i,t)$
&
Receiving flow rate of cell $(\ell,i)$,
expressed in $\mathrm{veh/h}$.
\\

$Y(\ell,i,t)$
&
Aggregate flow rate across the boundary from cell $(\ell,i)$ to
cell $(\ell,i+1)$, expressed in $\mathrm{veh/h}$.
\\

$F(\ell,i,h,t)$
&
Flow rate of commodity $h$ crossing boundary $i$ of link $\ell$,
expressed in $\mathrm{veh/h}$; $i=0$ and $i=I_{\ell}$ denote the
upstream and downstream link boundaries, respectively.
\\

\midrule
\multicolumn{2}{l}{\textbf{Traffic assignment and routing fractions}}\\

$\mathcal{K}(\ell,h)$
&
Set of admissible outgoing links from which destination $r(h)$
remains reachable.
\\

$p(\ell,k,h,t)$
&
Fraction of commodity $h$ routed from incoming link $\ell$ to
outgoing link $k$ at time $t$.
\\

\midrule
\multicolumn{2}{l}{\textbf{Signalized node model}}\\

$D(\ell,k,h,t)$
&
Movement flow demand of commodity $h$ for movement $(\ell,k)$,
expressed in $\mathrm{veh/h}$.
\\

$D(\ell,k,t)$
&
Aggregate flow demand for movement $(\ell,k)$,
expressed in $\mathrm{veh/h}$.
\\

$D^{m}(\ell,k,t)$
&
Flow demand of vehicle class $m$ for movement $(\ell,k)$,
expressed in $\mathrm{veh/h}$.
\\

$\alpha(\ell,k,m,t)$
&
Fraction of vehicle class $m$ in the demand for movement $(\ell,k)$.
\\

$\bar{\tau}^{\mathrm{mov}}(\ell,k,t)$
&
Composition weighted time gap associated with movement $(\ell,k)$.
\\

$\mu(\ell,k,t)$
&
Composition dependent saturation flow rate of movement $(\ell,k)$,
expressed in $\mathrm{veh/h}$.
\\

$g(\ell,k,t)$
&
Effective green duration available to movement $(\ell,k)$ during
simulation step $t$, expressed in $\mathrm{s}$.
\\

$\sigma(\ell,k,t)$
&
Binary signal indicator for a simulation step that is fully green or
fully red.
\\

$\hat f(\ell,k,t)$
&
Candidate aggregate movement flow rate before downstream supply
allocation, expressed in $\mathrm{veh/h}$.
\\

$f(\ell,k,t)$
&
Realized aggregate flow rate through movement $(\ell,k)$,
expressed in $\mathrm{veh/h}$.
\\

$f(\ell,k,h,t)$
&
Realized flow rate of commodity $h$ through movement $(\ell,k)$,
expressed in $\mathrm{veh/h}$.
\\

$z(\ell,h,t)$
&
Flow rate of commodity $h$ exiting the modeled network at its
destination, expressed in $\mathrm{veh/h}$.
\\

\midrule
\multicolumn{2}{l}{\textbf{External demand, virtual queues, and downstream supply}}\\

$\mathcal{L}^{\mathrm o}$
&
Set of links through which external demand enters the modeled network.
\\

$a(\ell,h,t)$
&
External arrival flow rate of commodity $h$ at origin link $\ell$,
expressed in $\mathrm{veh/h}$.
\\

$b(\ell,h,t)$
&
Number of vehicles of commodity $h$ waiting in the virtual origin
queue of link $\ell$ at time $t$.
\\

$\hat e(\ell,h,t)$
&
Candidate external entry flow rate of commodity $h$ into origin
link $\ell$, expressed in $\mathrm{veh/h}$.
\\

$e(\ell,h,t)$
&
Realized external entry flow rate of commodity $h$ into origin
link $\ell$, expressed in $\mathrm{veh/h}$.
\\

$\hat y^{\mathrm{in}}(k,t)$
&
Total candidate inflow rate into receiving link $k$,
expressed in $\mathrm{veh/h}$.
\\

$\chi(k,t)$
&
Proportional downstream supply allocation factor for receiving link $k$.
\\

\bottomrule
\end{tabular}

\end{minipage}

\end{table*}